\documentclass[11pt,a4paper]{article}
\pdfoutput=1

\usepackage{amssymb,amsmath,slashed,latexsym}
\usepackage{xcolor}
\usepackage{float}
\usepackage{footmisc}

\usepackage[hmarginratio=1:1,left=26mm,top=32mm,columnsep=20pt]{geometry} % Document margins

\usepackage{graphicx}
\usepackage[percent]{overpic}
\usepackage{caption}
\usepackage{subcaption}

\usepackage{tikz}
\usetikzlibrary{positioning,arrows}
\usetikzlibrary{decorations.pathmorphing}
\usetikzlibrary{decorations.markings}
\usetikzlibrary{calc}
\usetikzlibrary{matrix}
\usepackage{bbold}
\usepackage[vcentermath]{youngtab}

\usepackage[T1]{fontenc}
\usepackage{microtype} 

\usepackage{color}
\definecolor{dark-gray}{gray}{0.20}
\definecolor{gray}{gray}{0.30}
\definecolor{light-gray}{gray}{0.80}
\definecolor{dark-red}{rgb}{0.7,0,0}
\definecolor{dark-green}{rgb}{0.1,0.4,0}
\definecolor{dark-blue}{rgb}{0.3,0.3,0.7}
\definecolor{light-blue}{rgb}{0.8,0.8,1}

\usepackage{cite}
\usepackage{hyperref}
\hypersetup{
	colorlinks=true,
	linkcolor=dark-blue,
	citecolor=dark-red,
	urlcolor=dark-green,
	linktoc=page
}

\usepackage{titlesec} % Allows customization of titles
\titleformat{\section}[block]{\large\bfseries\centering}{\thesection}{1em}{} % Change the look of the section titles
\titleformat{\subsection}[block]{\bfseries}{\thesubsection}{1em}{} % Change the look of the section titles
\numberwithin{equation}{section}
\usepackage{abstract}

\def\Im{\mathop{\rm Im}\nolimits}
\def\trace{\mathop{\rm Tr}\nolimits}
\def\rme{{\rm e}}
\def\rmi{{\rm i}}

\newsavebox{\uuunit}
\sbox{\uuunit}
{\setlength{\unitlength}{0.825em}
	\begin{picture}(0.6,0.7)
	\thinlines
	\put(0,0){\line(1,0){0.5}}
	\put(0.15,0){\line(0,1){0.7}}
	\put(0.35,0){\line(0,1){0.8}}
	\multiput(0.3,0.8)(-0.04,-0.02){12}{\rule{0.5pt}{0.5pt}}
	\end {picture}}

\csname @addtoreset\endcsname{equation}{section}
\newcommand{\SU}{\mathop{\rm SU}}
\newcommand{\SO}{\mathop{\rm SO}}
\newcommand{\SL}{\mathop{\rm SL}}
\newcommand{\U}{\mathop{\rm {}U}}

\newcommand{\N}{\mathcal{N}}

\newcommand{\dd}{\mathrm{d}}

\newcommand{\e}{\mathrm{e}}
\newcommand{\w}{\wedge}

\newcommand{\be}{\begin{equation}}
\newcommand{\ee}{\end{equation}}
\newcommand{\bea}{\begin{eqnarray}}
\newcommand{\eea}{\end{eqnarray}}

\newcommand{\f}[2]{\frac{#1}{#2}}
\newcommand{\R}{\mathbf{R}}

\newcommand{\vol}{\text{vol}}

\renewcommand{\Im}{\text{Im}}

\title{\vspace{0mm}\fontsize{22pt}{10pt}\selectfont\textbf{A Holographic Kaleidoscope for $\mathcal{N}=1^{*}$}\vspace{10mm}}

\author{Nikolay Bobev, Fri\dh rik Freyr Gautason, \\[0.1cm]Benjamin E. Niehoff, and Jesse van Muiden \\[15mm] % Your name
\normalsize Instituut voor Theoretische Fysica, KU Leuven\\
\normalsize Celestijnenlaan 200D, B-3001 Leuven, Belgium\\[5mm]
\texttt{\small\href{mailto:nikolay.bobev@kuleuven.be}{\{nikolay.bobev}, \href{mailto:ffg@kuleuven.be}{ffg}, \href{mailto:ben.niehoff@kuleuven.be}{ben.niehoff}, \href{mailto:jesse.vanmuiden@kuleuven.be}{jesse.vanmuiden\}@kuleuven.be}}
}
\date{}

\begin{document}  
%----------------------------------------------------------------------------------------

\maketitle

%----------------------------------------------------------------------------------------
%%%			ABSTRACT
%----------------------------------------------------------------------------------------

\begin{abstract}
\noindent

We study in detail the recently-found family of asymptotically AdS$_5\times S^5$ type IIB supergravity solutions dual to the $\N=1^*$ SYM theory with equal masses. The backgrounds exhibit a naked singularity and are labelled by a dimensionless parameter, $\lambda$, which is interpreted as the ratio of the gaugino condensate and the mass in the dual field theory. When $|\lambda|<1$ we show that the naked singularity is due to a smeared distribution of polarized $(p,q)$ five-branes. For this range of parameters we study the nature of the singularity using probe strings and show that the dual line operators exhibit screening behavior. These features are in line with the physics anticipated in the work of Polchinski-Strassler.  For $|\lambda|=1$ the naked singularity has qualitatively different behavior which has no clear brane interpretation. We show that when $\lambda=1$ the singularity can be excised and replaced by a smooth Euclidean supergravity solution with an $S^4$ boundary.
\end{abstract}
\noindent 
\thispagestyle{empty}
\vfill
\newpage

\setcounter{tocdepth}{2}
\tableofcontents

 %%%%%%%%%%%%%%%%
 \section{Introduction}
 \label{sec:intro}
 %%%%%%%%%%%%%%%%

Studying four-dimensional quantum field theories with a mass gap using holography is bound to offer insights into their strong-coupling dynamics. This was appreciated in the early days of AdS/CFT and an intense effort to construct and study examples of the holographic duality in a non-conformal setting was undertaken. This program is under good technical control for models arising from string or M-theory which preserve a certain amount of supersymmetry. Nevertheless, it still remains challenging to construct explicit supergravity solutions dual to a four-dimensional QFT in a confining vacuum. 

Two well-studied examples in this context are the Klebanov-Strassler \cite{Klebanov:2000hb} and Maldacena-N\'u\~nez \cite{Maldacena:2000yy} backgrounds in type IIB supergravity. Both examples present analytic supergravity solutions which are dual to a non-conformal vacuum of a supersymmetric QFT and they have been used extensively to study the dynamics of the gauge theory. It is worth noting however that in both of these setups there are some exotic features. The Klebanov-Strassler solution is dual to an $\SU(N+M)\times \SU(N)$ $\mathcal{N}=1$ quiver gauge theory which undergoes an infinite cascade of Seiberg dualities. This is manifested in the supergravity dual by the absence of an asymptotically locally AdS$_5$ region. In addition, it was shown in  \cite{Aharony:2000pp,Gubser:2004qj} that the vacuum of the gauge theory is not massive due to the presence of massless glueballs. The Maldacena-N\'u\~nez background arises from D5-branes wrapped on an $S^2$ so as to preserve $\mathcal{N}=1$ supersymmetry in four dimensions. The supergravity solution however does not exhibit a separation between the ``QCD scale'' and the KK scale which is problematic for interpreting the holographic dual as a four-dimensional field theory.
 
Our goal here is to revisit another well-known setup for constructing a gravitational dual to a massive supersymmetric QFT in four-dimensions, namely the $\mathcal{N}=1^{*}$ mass deformation of $\mathcal{N}=4$ SYM \cite{Girardello:1999bd,Polchinski:2000uf,Pilch:2000fu}. This gauge theory has a rich set of supersymmetric vacua which have been studied extensively in the past, see for example \cite{Vafa:1994tf,Donagi:1995cf,Dorey:1999sj} for a field theory discussion and \cite{Dorey:2000fc,Polchinski:2000uf,Aharony:2000nt} for an analysis in a holographic context. Some of the supersymmetric vacua have a mass gap and can be studied quantitatively using various tools. In particular it is possible to compute the low-energy effective superpotential in the massive vacua using the S-duality of the parent $\mathcal{N}=4$ SYM theory \cite{Dorey:1999sj,Dorey:2000fc,Aharony:2000nt} or matrix model techniques \cite{Dijkgraaf:2002dh}. However, the physics of other supersymmetric vacua of the theory is not amenable to study with these methods and remains poorly understood. 

The gauge/gravity duality offers an alternative vantage point that may elucidate the gauge theory physics. The first problem in this context is to construct explicit supergravity solutions dual to the supersymmetric vacua of the gauge theory. There are at least two approaches to address this. One can use the five-dimensional maximal $\SO(6)$  gauged supergravity theory of \cite{Gunaydin:1984qu,Pernici:1985ju,Gunaydin:1985cu} to construct asymptotically AdS$_5$ supersymmetric domain wall solutions which implement, holographically, the RG flow from the $\mathcal{N}=4$ SYM theory to some of the vacua of $\mathcal{N}=1^*$. This was pursued in \cite{Girardello:1999bd} (see also \cite{Pilch:2000fu}), where explicit analytic supergravity solutions of this type were found. The GPPZ solutions in \cite{Girardello:1999bd} are dual to the $\mathcal{N}=1^{*}$ theory with equal values of the mass parameters and thus enjoy an $\SO(3)$ flavor symmetry. They exhibit a naked singularity in the IR, which prohibits the study of their physics using five-dimensional supergravity. An alternative approach is to implement the $\mathcal{N}=1^*$ mass deformation directly in type IIB supergravity by a suitable deformation of the AdS$_5\times S^5$ solution which is dual to the $\mathcal{N}=4$ conformal vacuum. The mass deformation of the gauge theory breaks the $\SO(6)$ R-symmetry of $\mathcal{N}=4$ SYM and thus one has to look for ten-dimensional supergravity solutions with little or no isometry on the internal $S^5$. This is clearly a technically challenging problem. Nevertheless, progress was made in this direction by using various approximations \cite{Polchinski:2000uf}. As shown in \cite{Polchinski:2000uf} the mass deformation in the gauge theory amounts to turning on R-R and NS-NS three-form flux on $S^5$. The D3-branes which make up the undeformed AdS$_5\times S^5$ background are affected by this flux and undergo polarization to five-branes through the Myers effect \cite{Myers:1999ps}. Compelling evidence for this polarization mechanism was presented in \cite{Polchinski:2000uf}, and a map between some of the supersymmetric vacua of $\mathcal{N}=1^*$ and polarized $(p,q)$ five-branes was proposed. Nevertheless, a fully backreacted supergravity solution which captures this physics remains out of reach. A possible way to remedy this impasse is to exploit the fact that the five-dimensional maximal supergravity is a consistent truncation of type IIB supergravity on $S^5$. This was suspected to be true for a long time but was rigorously established only recently in \cite{Lee:2014mla,Baguet:2015sma}. Using the explicit uplift formulae of \cite{Pilch:2000ue,Baguet:2015sma} one can find analytic ten-dimensional solutions which are the uplift of the GPPZ solutions. Recently this was done explicitly in \cite{Petrini:2018pjk,Bobev:2018eer}. 

The goal of our work is to study the naked singularity of the GPPZ solutions in ten dimensions, understand the physics of the vacuum in the dual gauge theory, and shed light on some of the qualitative features anticipated by Polchinski-Strassler. To this end we provide a brief summary of the ten-dimensional supergravity solutions of \cite{Petrini:2018pjk,Bobev:2018eer} and proceed to study their behavior near the naked singularity. The backgrounds in \cite{Petrini:2018pjk,Bobev:2018eer} are labelled by a real parameter $\lambda$ which is the holographic dual of the dimensionless ratio of the gaugino bilinear vev and the mass in the dual gauge theory. The criteria proposed in \cite{Gubser:2000nd,Maldacena:2000mw} for physically acceptable naked singularities in string theory restrict the value of $\lambda$ to lie in the range $-1\leq\lambda\leq 1$. For $|\lambda|<1$ we find that the naked singularity is smeared, in an $\SO(3)$ invariant way, along a one-dimensional submanifold of $S^5$ parametrized by an angular coordinate $\alpha$. The divergences of the ten-dimensional supergravity fields near this locus are compatible with those of a smeared distribution of polarized five-branes with an $\mathbf{R}^{1,3}\times S^2$ world-volume. This is qualitatively similar to the physics anticipated in \cite{Polchinski:2000uf} however there are some differences. We show that as one varies the angle $\alpha$ the supergravity background undergoes an $\SL(2,\mathbf{R})$ rotation. Therefore the type of polarized $(p,q)$ five-brane one finds near the singularity depends on the value of the angle $\alpha$. For example, at $\alpha=0$ we have an NS5-brane, while for $\alpha=\pi/4$ one finds an $(1,1)$ five-brane. A complementary way to understand the physics of the naked singularity is to study probe strings in the ten-dimensional background. We perform a detailed analysis of $(m,n)$ probe string solutions for $|\lambda|<1$  and find additional evidence for the interpretation of the naked singularity as a smeared distribution of polarized $(p,q)$ five-branes. The regularized on-shell action of these probe strings is dual to the expectation value of line/loop operators in the $\mathcal{N}=1^{*}$ gauge theory. Our calculations show that the vevs of the loop operators exhibit a screening behavior. This suggests that the supergravity solutions with $|\lambda|<1$ are not dual to a confining vacuum of $\mathcal{N}=1^{*}$.

The nature of the singularity in the solutions of \cite{Petrini:2018pjk,Bobev:2018eer} for $|\lambda|=1$ is qualitatively different and does not admit an interpretation as polarized five-branes. We study probe D3-branes and show that for $|\lambda|=1$ they have vanishing effective tension near the singularity. This signals the presence of new light modes near the singularity and suggests that one should not interpret the singular solution in supergravity. We show explicitly how to regulate the singular supergravity solution with $\lambda=1$ while preserving supersymmetry. To do this one has to employ the regular Euclidean supergravity solutions in \cite{Bobev:2016nua}. These solutions are dual to the $\mathcal{N}=1^{*}$ theory on $S^4$ of radius $\mathcal{R}$ and we show that for large values of $\mathcal{R}$ one finds $\lambda=1$. The existence of these smooth solutions with an $S^4$ boundary suggests that for $\lambda=1$ the planar $\mathcal{N}=1^{*}$ theory is in a massive vacuum.

The $\mathcal{N}=1^{*}$ gauge theory admits supersymmetric vacua with non-vanishing vevs for bosonic bilinear operators in the $\mathbf{20}'$ of $\SO(6)$. The five-dimensional gauged supergravity truncation we use contains a scalar dual to one of these operators and we look for supersymmetric domain wall solutions with nontrivial vevs for it. We find that there are no such supersymmetric solutions which are physically acceptable according to the criteria in \cite{Gubser:2000nd,Maldacena:2000mw}.

In the next section we present a short summary of well-known results about the $\mathcal{N}=1^*$ SYM theory and its vacuum structure. We also briefly discuss the Polchinski-Strassler description of some of the gauge theory vacua in terms of polarized five-branes. In Section~\ref{sec:5Dsugra} we show how to construct the GPPZ solution in a consistent truncation of five-dimensional supergravity and show that there are no other physically relevant solutions in this truncation. In Section~\ref{sec:10Dsolution}  we analyze in detail the ten-dimensional uplift of the GPPZ solution and the nature of the naked singularity. To this end we study probe strings and D3-branes in the ten-dimensional background and their dual gauge theory interpretation. We conclude in Section~\ref{sec:discussion} with a discussion on the implications of our results for holography and some of the open problems. The four appendices contain an amalgam of technical results used in the main text.

%%%%%%%%%%%%%%%%
\section{The $\N=1^*$ field theory}
\label{Sec:FieldTheory}
%%%%%%%%%%%%%%%%

The $\N=1^*$ theory is a deformation of the $\N=4$ SYM theory. To establish our notation we start with a brief review of $\N=4$ SYM. 

The ${\cal N}=4$ vector multiplet consists of a gauge field\footnote{In this paper we choose the gauge group to be $\SU(N)$.} $A_\mu$, four gaugini $\psi^m$, and six scalars $X_I$, all of which transform in the adjoint of the gauge group $\SU(N)$. The ${\cal N}=4$ SYM enjoys a $\SU(2,2|4)$ superconformal symmetry. The bosonic subalgebra consists of the $\SU(2,2)\simeq \SO(2,4)$ four-dimensional conformal algebra and an $\SU(4)\simeq \SO(6)$ R-symmetry. The fermions transform in the ${\bf 4}$ of $\SU(4)$, the scalars transform in the ${\bf 6}$ and the vector is a singlet. The Lagrangian can be written as\footnote{The fermions $\psi^m$ are four-dimensional left-handed Majorana spinors whereas $\psi_m$ are right-handed.}
\begin{equation}\label{SYMLag}
\mathcal L = \frac{1}{g_\text{YM}^2}  \trace \Big( \frac12 |F|^2 + |D X_I|^2 +  \overline \psi_m \slashed{D} \psi^m + (\overline{\psi}^m \left[X_{mn},\psi^n\right]+\text{h.c.}) +  \left[X_I,X_J\right]^2 \Big)+ \frac{\theta}{8\pi^2} F\wedge F ~,
\end{equation}
where to write the Yukawa interaction terms we have transformed the $\SO(6)$ index $I$ to a pair of antisymmetric $\SU(4)$ indices $mn$. In this paper we focus on a mass deformation of the $\N=4$ theory that preserves $\N=1$ supersymmetry. It is therefore convenient to write the $\N=4$ theory in manifestly $\N=1$ language. This is achieved by organizing the $\N=4$ vector multiplet into an $\N=1$ vector multiplet, $V$, and three chiral multiplets, $\Phi_i$, as follows
\begin{equation}
V = \left(A_{\mu} , \psi_4\right),\quad \Phi_i = \left(\psi_i,\phi_i\right),
\end{equation}
where $\phi_i = \left(X_i + \rmi X_{i+3}\right)/\sqrt{2}$. In this rewriting of the theory only an $\SU(3)\times\U(1)_r\subset \SU(4)$ R-symmetry is manifest. The index $i=1,2,3$ transforms in the fundamental representation of $\SU(3)$. From the perspective of $\N=1$ supersymmetry, the global $\SU(3)$ symmetry  can be viewed as a flavor symmetry. The advantage of writing $\N=4$ SYM in $\N=1$ language is that the chiral Lagrangian is fully determined by the K\"{a}hler and super potentials
\begin{equation}
K = \frac{1}{g_\text{YM}^2}\trace \, \Phi_i^\dagger \Phi_i\,,\qquad W = \f{1}{g_\text{YM}^2} \trace \, \left[\Phi_1, \Phi_2\right]\Phi_3\,.
\end{equation}
It is now easy to write down the mass deformation of interest in this work as the following modification of the superpotential above
\begin{equation}\label{massdef}
\Delta W = \f{1}{g_\text{YM}^2}\trace \left( m_1 \Phi_1^2 + m_2 \Phi_2^2+ m_3 \Phi_3^2\right)\,.
\end{equation}
Here $m_{1,2,3}$ are three independent complex parameters. For generic choices of $m_{1,2,3}$ supersymmetry is explicitly broken to $\N=1$, however for the specific choice $m_1=m_2\ne 0$ and $m_3=0$, the two chiral multiplets $\Phi_1$ and $\Phi_2$ combine into an $\N=2$ hyper multiplet and the $\mathcal{N}=1$ vector multiplet together with $\Phi_3$ form an $\N=2$ vector multiplet and we obtain the so called $\N=2^*$ theory. The Lagrangian then enjoys $\N=2$ supersymmetry where the $\SU(3)$ symmetry is broken to $\SU(2)_R\times \U(1)$. The $\N=2$ R-symmetry is a product of $\SU(2)_R$ and a linear combination of $\U(1)$ and $\U(1)_r$. Another special deformation is obtained by setting two of the masses to zero. In this case the theory flows to an interacting conformal fixed point in the IR \cite{Leigh:1995ep,Khavaev:1998fb,Freedman:1999gk}. In this paper we focus on the deformation in \eqref{massdef} where we take the three masses equal, i.e. $m=m_1=m_2=m_3$. In this case the $\SU(3)$ flavor symmetry is broken to its real subalgebra $\SO(3)$. This theory exhibits a rich vacuum structure which was studied in \cite{Donagi:1995cf}, and discussed further in \cite{Polchinski:2000uf}. 

%%%%%%%%%%%%%%%%%%%%%%%
\subsection{Vacua of $\N=1^*$}
%%%%%%%%%%%%%%%%%%%%%%%

The classical vacua are determined by solving the F-term equation
\be\label{classicalVac}
[\phi_i,\phi_j ] = -m\varepsilon_{ijk} \phi_k\,.
\ee
Since all matter fields are in the adjoint representation of $\SU(N)$, the solutions to these equations are given by $N$-dimensional representations of $\SU(2)$. A generic $\SU(2)$ representation is of course reducible and therefore a vacuum of the theory is determined by a partition of $N$, such that
\begin{equation}\label{eq:repsum}
\sum \limits_{d=1}^N d \, k_d = N\,.
\end{equation}
Here $k_d$ are non-negative integers that determine the frequency of the appearance of the $d$-dimensional irreducible representation of $\SU(2)$. Almost all classical vacua break the $\SU(N)$ gauge group and the preserved gauge symmetry is $(\prod_d \U(k_d))/\U(1)$. Note that for any divisor $D$ of $N$ (including $N$ itself) the vacuum specified by taking $k_D=N/D$ and all other $k_d=0$ has a preserved gauge group $\SU(N/D)$. As we discuss below, it is justified to refer to these as the massive vacua of the theory. The case $D=N$ is distinguished as the \emph{classical} massive vacuum with a completely broken gauge group and is called the \emph{Higgs} vacuum. A  solution to the classical vacuum equations \eqref{classicalVac} in which multiple distinct $k_d$ are nonzero (thus, not falling into the class of massive vacua just discussed) will have at least one unbroken $\U(1)$ gauge group factor and is therefore a  \emph{Coulomb} vacuum.

Quantum mechanically, the structure is quite a bit richer.  As demonstrated in \cite{Donagi:1995cf}, the massive vacua (with an unbroken $\SU(N/D)$ gauge group) split into $N/D$ separate vacua, which can be classified using the algebra of line/loop operators developed in \cite{tHooft:1977nqb}.  To each such operator one associates a pair of integers $x = (m,n)$ which represent its electric and magnetic charges.\footnote{These operators can be thought of as products of $m$ Wilson line operators and $n$ 't~Hooft line operators.} These charges take values in the compact charge lattice $F = \mathbf{Z}_N^m\times \mathbf{Z}_N^e$, where $\mathbf{Z}_N^e$ is the center the gauge group and $\mathbf{Z}_N^m = \pi_1 \left[ \SU(N)/\mathbf{Z}_N^e \right]$. The algebra of loop operators is equipped with a natural pairing $\langle\cdot,\cdot\rangle: F\to \mathbf{Z}_N$ such that for $x=(m,n)$ and $y=(m',n')$
\be
\langle x,y\rangle = m n' - m' n \quad\text{mod}~N\,.
\ee
In direct analogy with the Meissner effect, the condensation of a charge $x$ leads to a confinement of any charge $y$ for which $\langle x,y\rangle \ne0$. Furthermore, two charges $x$ and $z$ which simultaneously condense have zero product: $\langle x,z\rangle =0$. One can then deduce (as was shown in \cite{tHooft:1977nqb}) that the vacua with a mass gap are precisely those for which $N$ charges (electric or magnetic) condense and all others confine.  These vacua correspond 1-to-1 with the $N$-dimensional subgroups of $F = \mathbf{Z}_N^m\times \mathbf{Z}_N^e$. Any such subgroup can be generated by a pair of elements \cite{Donagi:1995cf}
\be \label{xy generators}
x=(b,D)~,\qquad y=(N/D,0)~, \qquad \mod N~,
\ee
where $D$ is a positive divisor of $N$ and $0\le b \le N/D-1$. The elements of each such subgroup then label the charges that condense in that vacuum.  The classical Higgs vacuum with completely broken gauge group has $N/D = 1$ and therefore has a unique quantum representative generated by $x = (0,0)$ and $y = (1,0)$.  The classical vacuum with $d=1$, $k_d=N$ has $N/D = N$ and therefore splits into $N$ quantum vacua, the \emph{confining} vacua, generated by $x = (0,1)$ through $x = (N-1,1)$ and $y = (0,0)$.  In between are the vacua with $N/D = k_D$ for some integer $1 < k_D < N$; these are the \emph{oblique confining} vacua where some mixture of electric and magnetic charges condense, and to each such classical vacuum there correspond $k_D$ massive, quantum vacua.  The $\SL(2,\mathbf{Z})$ electric-magnetic duality acts on the charge lattice as follows \cite{Montonen:1977sn}:
\be
\begin{split}
T :~&(m,n) \mapsto (m+n,n)\quad\text{mod}~N\,,\\
S:~&(m,n)\mapsto (-n,m)\quad\text{mod}~N\,,\\
\end{split}
\ee
which induces a non-trivial duality between the massive vacua. In particular the Higgs vacuum is invariant under $T$, but under $S$ gets mapped to the $b = 0$ confining vacuum generated by $(0,1)$.  The confining vacua are permuted by $T$, which has the effect of incrementing the parameter $b \to b + 1 \mod N$.  Under $S$, the $(0,1)$ confining vacuum is mapped to the Higgs vacuum, whereas the other $(b,1)$ confining vacua are mapped to various oblique vacua.  In general, $T$ will permute (by varying $b$) the quantum vacua corresponding to a given classical vacuum (with particular $D$), whereas $S$ will act in a way that exchanges different classical vacua.  In the special case where $N = D^2$ is a square number, then the vacuum generated by  
\be\label{selfdualvac}
x=(0,D)~,\qquad y=(D,0)\,,
\ee
is $S$-duality invariant. This is the only massive vacuum invariant under $S$-duality and we refer to it as the \emph{self-dual} vacuum.

The infrared physics of the field theory in one of the massive vacua is controlled by the effective superpotential.  This was computed in \cite{Dorey:1999sj} for a subclass of the massive vacua by reducing the field theory on a circle to three dimensions and identifying an integrable system that controls the dynamics. This result was later extended in \cite{Dorey:2000fc,Polchinski:2000uf,Aharony:2000nt} to include all the massive vacua and arrive at the following IR effective superpotential 
\be\label{IRsuperpot}
W_\text{IR} = \f{m_1m_2m_3 N^2}{24}\left[E_2(\tau) - \f{N}{D^2}E_2\left(\f{N\tau}{D^2}+\f{b}{D}\right) + A(\tau,N)\right]\,.
\ee
Here the three masses $m_i$ are generic and $\tau = 4\pi \rmi/g_\text{YM}^2 +\theta/2\pi$. The function $E_2(\tau)$ is the regulated Eisenstein series of modular weight two and $A(\tau,N)$ is an undetermined holomorphic function of $\tau$ discussed in some detail in \cite{Dorey:2000fc,Aharony:2000nt}.  Using the superpotential we can compute the chiral and gluino condensates\footnote{The operator $\trace (\psi_4\psi_4+2\Phi_1[\Phi_2,\Phi_3])$ receives a correction at one loop due to the Konishi anomaly which we are ignoring.} 
\be\label{eq:chgvev}
\left\langle \trace \Phi_i^2\right\rangle = g_\text{YM}^{2}\f{\partial W_\text{IR}}{\partial m_i}~,\qquad \left\langle\trace (\psi_4\psi_4+2\Phi_1[\Phi_2,\Phi_3])\right\rangle = -16\pi \rmi\left(\f{\partial W_\text{IR}}{\partial \tau}-\rmi\f{ W_\text{IR}}{\Im\tau}\right)\,.
\ee
We note that in the self-dual vacuum \eqref{selfdualvac}, the Eisenstein terms vanish and the superpotential reduces simply to the holomorphic function $A(\tau,N)$.  Thus the chiral condensate $\left\langle \trace \Phi_i^2\right\rangle$ is proportional to $A(\tau,N)$. We also note that there are subtle questions about operator mixing along the RG flow from the $\mathcal{N}=4$ theory to a given vacuum of $\mathcal{N}=1^{*}$. These were discussed in \cite{Aharony:2000nt} but a fully general analysis is not present in the literature.

So far, we have focused on the massive vacua; however, the vast majority of vacua have unbroken $\U(1)$ gauge factors, and therefore massless photons in the IR. These Coulomb vacua do not yet have an elegant classification in the literature along the lines given for the massive vacua. As we explain below, both massive and Coulomb vacua will play a role in interpreting our holographic solutions. In \cite{Polchinski:2000uf} some properties of the Coulomb vacua were determined using inspiration from the physics of five-branes in type IIB string theory which we now review. 

%%%%%%%%%%%%%%%%%%%%%%%%%%
\subsection{Relation to five-branes}
%%%%%%%%%%%%%%%%%%%%%%%%%%

Polchinski and Strassler argued that the vacua of $\N=1^*$ are related to the polarization of D3-branes into five-branes which are immersed in a three-form flux background of IIB string theory \cite{Polchinski:2000uf}. The strength and shape of this three-form flux is controlled by the three mass parameters of the field theory. The physics of polarized branes studied by Myers in \cite{Myers:1999ps} shows features reminiscent of the discussion of the $\N=1^*$ vacua above. For completeness we sketch the arguments in \cite{Polchinski:2000uf} that lead to the mapping of the massive vacua of $\N=1^*$ to the polarization states of three-branes and their five-brane interpretation.

Consider a stack of D3-branes in a constant background RR three-form flux background.  It proves convenient to dualize the three-form to a seven-form and write it in terms of a six-form potential\footnote{Here we treat the branes as probes and assume that the dilaton is constant and $F_5$ vanishes in the background.}
\be
 F_7 =  -g_s^{-1}\star_{10}F_3 = \dd C_6\,.
\ee
The effective action for a stack of D3-branes contains couplings of the form
\be\label{D3couplings}
\mu_3 \int \trace P \left[C_4 +2\pi \rmi\ell_s^2 \iota_{X}\iota_X C_6\right]\,,
\ee
where $P$ denotes the pull-back of the ten-dimensional fields onto the brane world-volume, $\mu_3$ is the charge of the D3-branes and $X = X^I\partial_I$ denotes collectively the coordinates transverse to the D3-branes. Since we are dealing with a stack of D3-branes the transverse coordinates are now matrix-valued and transform in the adjoint of the gauge group living on the brane. The appearance of $C_6$ in this action shows that, for a non-abelian configuration of the $X$'s, the D3-branes carry a D5-brane charge. For non-abelian D-branes Myers argued that the DBI term is modified to include commutators of $X$. For static D3-branes in flat space the DBI action reduces to the potential of $\N=4$ SYM given in \eqref{SYMLag}
\be
V_\text{DBI} = \mu_3N  + \mu_3\pi^2\ell_s^4\trace \left[X^I,X^J\right]^2+\cdots\,,
\ee
where the dots stand for corrections obtained by expanding the square root in the DBI action to higher order in the coordinates $X^I$. Let us assume that the D3-branes extend along the coordinates $x^{0,4,5,6}$ and the three-form $F_3$ is constant in the three transverse directions $x^{7,8,9}$. Then the seven-form can be written as $F_7 = f \epsilon_{ijk}\vol_4 \w\dd x^i\w\dd x^j\w\dd x^k$ where $i,j,k=1,2,3$ and $f$ determines the magnitude of the flux. Minimizing the probe action we find the vacuum equation for the stack of D3-branes\cite{Myers:1999ps}
\be\label{polbranes}
[X_i,X_j] = \rmi f\epsilon_{ijk}X_k\,,
\ee
which has a form similar to the classical $\N=1^*$ vacuum equation \eqref{classicalVac}. Equation \eqref{polbranes} shows that the flux induces a polarization of the D3 branes, and they arrange themselves on a (fuzzy) two-sphere of radius proportional to $f$ \cite{Myers:1999ps}. This fuzzy sphere carries D5-brane charge according to \eqref{D3couplings} and therefore has a dual interpretation in terms of D5-branes. The D5-brane charge of the polarized state depends on which solution of \eqref{polbranes} is realized. The lowest energy solution is the irreducible one that corresponds to unit D5-brane charge. Other irreducible representations correspond to polarization to multiple two-spheres, each carrying their own D5 charge. By relating the magnitude of the three-form flux $f$ to the mass deformation of the gauge theory, this argument shows that the physics of polarized D3-branes should play an important role in the holographic description of the $\N=1^*$ gauge theory.  The same conclusion can be reached for D3-branes in NS-NS three-form background where now the radius of polarization is scaled by a factor of $g_s^{-1}$. Myers showed that there exists a dual description in terms of a single stack of spherical D5-branes with $N$ units of D3-brane charge encoded in the flux of its world-volume $\U(1)$ gauge field. Again this can be understood by studying the coupling of the D5-brane to the ten-dimensional R-R potentials
\be
\mu_5 \int P\left[C_6 + 2\pi\ell_s^2 {\cal F}\w C_4\right]\,,
\ee
where $2\pi\ell_s^2 {\cal F} = 2\pi\ell_s^2 F+ P[B_2]$ and $F$ is the world-volume $\U(1)$ field strength. Let us now consider the gauge flux $F = (N/2) \vol_2$, where $\vol_2$ is the volume form of the two-sphere in the $(x^1,x^2,x^3)$ plane in polar coordinates. The normalization is chosen such that the quantized flux of ${\cal F}$ equals $N$ ($B_2$ is assumed to vanish). One can show that the static configuration for a D5-brane with such a world-volume flux in the above background is $\R^{1,3}\times S^2$ where the radius of the sphere matches the non-commutative picture above. In the foregoing discussion we assumed that the D3-branes would polarize into a single stack of D5-branes. However, the situation can be more involved. For example, the D3-branes could polarize into $D$ D5-branes each carrying $N/D$ D3-brane charge. This can be further generalized to polarization into $(p,q)$ five-branes, with NS5-brane charge $p$ and D5-brane charge $q$.

It is reasonable to expect that the holographic description of the vacua of $\N=1^*$ involves polarized five-branes of various flavors. Indeed Polchinski and Strassler found non-trivial evidence that this expectation is realized \cite{Polchinski:2000uf}. They constructed an approximate solution to type IIB supergravity by deforming AdS$_5\times S^5$ with 3-form fields in a small mass (flux) expansion which asymptote to $(p,q)$ five-branes in the IR. In particular they argued that the Higgs vacuum should correspond to a single polarized D5-brane, and the confining vacuum to a single polarized NS5-brane. The various oblique confining vacua are then described in terms of polarized $(p,q)$ five-branes. This correspondence between massive vacua of $\N=1^*$ and polarized five-brane states in type IIB string theory is supported by the fact that the $\SL(2,\mathbf{Z})$ of the gauge theory and the $\SL(2,\mathbf{Z})$ of type IIB string theory act identically on the vacua and the five-branes. In \cite{Polchinski:2000uf} it was also argued that the Coulomb vacua are dual to multiple stacks of $(p,q)$ five-branes of different radii. This is inspired by a direct analogy between the solutions of \eqref{polbranes} and \eqref{classicalVac}.

%%%%%%%%%%%%%%%%
\section{Five-dimensional supergravity}
\label{sec:5Dsugra}
%%%%%%%%%%%%%%%%

The supergravity dual of (at least some vacua of) the $\N=1^*$ theory can be constructed using five-dimensional $\mathcal{N}=8$ $\SO(6)$ gauged supergravity \cite{Girardello:1999bd,Pilch:2000fu}. When all three masses are equal, one can use the $\SO(3)$ flavor symmetry of the model to restrict to the corresponding $\SO(3)$-invariant subsector of the five-dimensional $\mathcal{N}=8$ theory. This was discussed in detail in \cite{Pilch:2000fu} where it was found that the $\SO(3)$-invariant truncation contains eight real scalar fields in addition to the five-dimensional metric. This consistent truncation can be truncated further by imposing additional discrete symmetries. A particular choice of discrete group leads to a supergravity model with four real scalars \cite{Pilch:2000fu,Bobev:2016nua}.  In the dual field theory, imposing this discrete symmetry corresponds to restricting the complex mass parameter and gaugino bilinear vev to be real. The solution of GPPZ \cite{Girardello:1999bd} is a particular solution of this five-dimensional model for which only two of the four scalars flow.\footnote{In \cite{Bobev:2016nua} it was shown that the four-scalar model, with all four scalars developing non-trivial profiles, allows for a family of solutions dual to the equal mass $\N=1^*$ theory on $S^4$. We discuss these solutions further below.} In this section we will briefly review the four-scalar model and its solutions. We refer to \cite{Bobev:2016nua} for further discussion on it.

The five-dimensional Lagrangian\footnote{In contrast to \cite{Bobev:2018eer}, we will work entirely in mostly-plus signature.} can be written in terms of the metric and two complex scalars $z^i$:
\begin{equation}
\mathcal L = \f{1}{4\pi G_N}\sqrt{\left|g\right|}\left(\frac14 R +\f12 \mathcal K_{i\bar \jmath} \partial_\mu z^i \partial^\mu \bar z ^{\bar \jmath} - \mathcal P \right)\,,
\end{equation}
where the K\"{a}hler potential, $\mathcal{K}$, determines the kinetic term metric $\mathcal K_{i\bar \jmath}$, and the superpotential $\mathcal{W}$ specifies the scalar potential $\mathcal{P}$ via the relations
\begin{equation}\label{Eq: 5D lagrangian}
\begin{aligned}
\mathcal K_{i\bar \jmath} =& \,\partial_i \partial_{\bar j} \mathcal K\,, \hspace{3.3cm} \mathcal K = - \log \left[ \left(1-z_1 \bar z_1\right)\left( 1-z_2 \bar z_2 \right)^3 \right]\,,\\
\mathcal W =& \frac{3 g}{4} \left(1+z_1 z_2\right)\left( 1-z_2^2 \right)\,,\quad \mathcal P = \frac{1}{2} \rme^{\mathcal K} \left[\mathcal K^{i\bar \jmath} D_i \mathcal W D_{\bar \jmath} \overline{\mathcal{W}} - \frac83 \mathcal W \overline{\mathcal W}\right]\,.
\end{aligned}
\end{equation}
Here $g$ is the gauge coupling constant of the parent $\N=8$ supergravity theory and the K\"{a}hler covariant derivative is defined as $D_i f= \left(\partial_i + \partial_i \mathcal K\right) f$. This model admits supersymmetric domain wall solutions with metric
\be\label{eq:5dmetansatz}
\dd s_5^2 = \dd r^2 + \e^{2A}\dd s_4^2~,
\ee
where $\dd s_4^2$ denotes the flat metric on Minkowski space.  The metric function $A$ and the scalars $z_{1,2}$ are assumed to only depend on the radial coordinate $r$. The BPS equations of the model are obtained by imposing that part of the supersymmetry of the ${\cal N}=8$ supergravity theory is preserved, i.e. by demanding that the fermion supersymmetry variations $\delta\psi_\mu$ and $\delta \chi$ vanish. The BPS equations take the form 
\begin{equation}\label{5DBPSflat}
{\cal E}_A \equiv A' - \frac23 \rme^{{\mathcal K}/2}|\mathcal W|=0\,,\qquad{\cal E}^i \equiv  (z^i)'  + \e^{{\mathcal K}/2} \f{{\mathcal W}}{|\mathcal W|} \mathcal K ^{i\bar \jmath} D_{\bar \jmath} \overline{\mathcal W}=0\,.
\end{equation}
In these equations the prime denotes a derivative with respect to the radial coordinate $r$. A field configuration that solves the equations in \eqref{5DBPSflat} automatically provides a solution to the full set of equations of motion derived from the Lagrangian in \eqref{Eq: 5D lagrangian}. This can be readily seen by  rewriting the Lagrangian in \eqref{Eq: 5D lagrangian} supplemented with the Gibbons-Hawking boundary term as a sum of squares
\be\label{sumofsquares}
{\cal L} +\mathcal{L}_{\text GH} = \f{1}{4\pi G_N}\sqrt{|g|}\left[3{\cal E}_A^2 -\f12 {\cal E}^i{\cal K}_{i\bar \jmath}\overline{\cal E}^{\bar\jmath}\right]  + \f{1}{4\pi G_N}\partial_r\left(\sqrt{|g|} \rme^{{\mathcal K}/2}|\mathcal W|\right)~.
\ee
A simple solution of the BPS equations in \eqref{5DBPSflat} is given by the maximally supersymmetric AdS$_5$ vacuum which takes the form
\begin{equation}\label{SUSYAdS}
z_1 = z_2 = 0 \quad \text{and} \quad  A=\frac{gr}{2}\,.
\end{equation}
It is clear that the gauge coupling of the supergravity theory is related to the length scale of AdS$_5$ by $L=2/g$.

%%%%%%%%%%%%%%%%%%%%%%%%%%
\subsection{UV expansion and holographic renormalization}
%%%%%%%%%%%%%%%%%%%%%%%%%%

The domain wall solutions we are interested in are asymptotic to the AdS$_5$ solution in \eqref{SUSYAdS}. They realize, holographically, the RG flow triggered by the mass terms in \eqref{massdef}, and the asymptotically-AdS$_5$ region of the solution corresponds to the $\mathcal{N}=4$ UV conformal fixed point. We can solve the BPS equations \eqref{5DBPSflat} in a systematic expansion around the AdS vacuum and use holographic renormalization\footnote{See \cite{Skenderis:2002wp} for a review.} to map this solution to observables in the dual QFT. As in \cite{Bobev:2016nua}, it is convenient to perform this UV expansion after doing the following change of variables for the scalar fields 
\be\label{fieldredefinition}\begin{split}
z_1&= \tanh\f12\left(3\alpha+\varphi-3\rmi\phi+\rmi\phi_4\right)\,,\\
z_2&=\tanh\f12\left(\alpha-\varphi-\rmi\phi-\rmi\phi_4\right)\,.
\end{split}
\ee
In these variables the scalar potential takes the form
\be\label{eq:Potphi}
{\cal P} = -\frac{3g^2}{64}   \f{4 \cosh 4 \alpha~ \cos 2 (\phi+\phi_4)+7
   \cos (2 \phi-2 \phi_4)+\cos (6 \phi + 2\phi_4)+4 \cos 4 \phi}{\cos (3 \phi-\phi_4) \cos^3(\phi+\phi_4)}\,.
\ee
Notice the the potential is independent of the field $\varphi$. This implies that the BPS equations in \eqref{5DBPSflat} have an integral of motion. The BPS equations are rewritten in terms of the new variables in Appendix \ref{5DBPSeq}. 

Expanding the scalar potential \eqref{eq:Potphi} to quadratic order around the AdS$_5$ vacuum leads to the following masses for the four scalar fields
\be
m_{\phi_4}^2L^2 = -3~,\qquad m_\phi^2L^2 = -3~,\qquad m_\alpha^2 L^2 = -4~,\qquad m_\varphi^2 L^2 = 0~.
\ee
This indicates that the scalars $\phi$ and $\phi_4$ are dual to dimension $\Delta=3$ fermion bilinear operators, $\alpha$ is dual to a dimension $\Delta=2$ scalar bilinear and $\varphi$ is dual to a marginal operator. More precisely we have the following map between the bulk scalar fields and operators in $\mathcal{N}=4$ SYM\footnote{Note that the four-scalar model we use here is different from the one employed in \cite{Petrini:2018pjk}, where all four scalars are in the $\mathbf{10}\oplus\overline{\mathbf{10}}$ and correspond to the complexified operators on the first line of \eqref{eq:holomapscalars}.}
\begin{equation}\label{eq:holomapscalars}
\begin{split}
\phi \leftrightarrow & ~\mathcal{O}_{\phi}=\sum_{j=1}^{3} \text{Tr}(\psi_j \psi_j+\bar{\psi}_j\bar{\psi}_j)\,, \qquad \phi_4 \leftrightarrow ~\mathcal{O}_{\phi_4}=\text{Tr}(\psi_4 \psi_4+\bar{\psi}_4\bar{\psi}_4)\,,\\
\alpha \leftrightarrow  &~\mathcal{O}_{\alpha}=\sum_{j=1}^{3} \text{Tr}(\phi_j^2 +\bar{\phi}_j^2)\,, \qquad\qquad \varphi \leftrightarrow ~\mathcal{O}_{\varphi}=\text{Tr} \, F_{\mu\nu}F^{\mu\nu}\,.
\end{split}
\end{equation}

The UV expansion of the BPS equations has the following form
\be\label{UVexpflat}\begin{split}
\phi &= \hat{m} \epsilon^{1/2}-\frac{5}{6}\hat{m}^3\epsilon^{3/2}+ {\cal O}(\epsilon^{2})\,,\\
\phi_4 &= w \epsilon^{3/2}+ {\cal O}(\epsilon^2)\,,\\
\alpha &= v  \epsilon+ {\cal O}(\epsilon^2)\,,\\
\varphi &= \varphi_0 +  {\cal O}(\epsilon^2)\,,\\
A &= -\f{1}{2}\log\epsilon - \f{\hat{m}^2}{2} \epsilon + {\cal O}(\epsilon^2)\,.
\end{split}\ee
Here $\epsilon = \e^{-g r}$ is a small parameter controlling the distance from the AdS$_5$ boundary. The parameter $\hat{m}$ corresponds to a source term for the fermion bilinear operator in \eqref{eq:holomapscalars} and is proportional to the mass parameter in the $\mathcal{N}=1^*$ Lagrangian \eqref{massdef}.\footnote{There is a also a source for the Konishi operator $\sum_{j=1}^{3}\text{Tr}(\phi_j\bar{\phi}_j)$ but this operator does not correspond to a supergravity mode.}
The parameter $w$ is related to a vev for the gaugino bilinear operator in \eqref{eq:holomapscalars} and $v$ is related to a vev for the bosonic bilinear in \eqref{eq:holomapscalars}. We refer to these vev as the gaugino and chiral condensate, respectively. To compute the exact values of these vevs one must carefully perform the holographic renormalization procedure for the four-scalar model.

As we discuss in Section \ref{subsec:chiralcond} the only physically interesting flat-sliced domain wall solution is the GPPZ solution. We therefore restrict our holographic renormalization analysis to it. This analysis has already appeared in several places in the literature, see Section 5 of \cite{Skenderis:2002wp} as well as \cite{Bianchi:2001kw,Bianchi:2001de}. For the GPPZ flow one has $\alpha=\varphi=0$ and it proves useful to simplify the supergravity Lagrangian by using the scalar variables $m_{\rm GPPZ}$ and $\sigma_{\rm GPPZ}$ employed in \cite{Girardello:1999bd} since they have canonical kinetic terms. This is achieved by the following change of variables
\begin{equation}\label{eq:phimsig}
\begin{split}
\tan(\tfrac{1}{2}(\phi_4-3\phi)) &= -\tanh(\tfrac{1}{2}(\sqrt{3}m_{\rm GPPZ}-\sigma_{\rm GPPZ}))\,,\\
\tan(\tfrac{1}{2}(\phi_4+\phi)) &= \tanh(\tfrac{1}{6}(\sqrt{3}m_{\rm GPPZ}+3\sigma_{\rm GPPZ}))\,.
\end{split}
\end{equation}
The scalars $m_{\rm GPPZ}$ and $\sigma_{\rm GPPZ}$ have the following linearized expansion in the AdS$_5$ UV region
\begin{equation}
\begin{split}
m_{\rm GPPZ} &= \sqrt{3}\hat{m} \epsilon^{1/2} + \frac{\hat{m}^3}{\sqrt{3}}\epsilon^{3/2}+{\cal O}(\epsilon^{2})\,,\\
\sigma_{\rm GPPZ} &= (w-\hat{m}^3)\epsilon^{3/2} +{\cal O}(\epsilon^{2})\,,
\end{split}
\end{equation}
where we have used the same coefficients as in the asymptotic expansion in \eqref{UVexpflat}. With this at hand we can readily apply the results in Section 5 of \cite{Skenderis:2002wp} to find that the vev of the gaugino bilinear is given by\footnote{Note that this expression corrects a typo in Equation (4.34) of \cite{Bobev:2016nua}.} 
\begin{equation}\label{eq:gauginovevhol}
\langle{\cal O}_{\phi_4} \rangle = \frac{N^2}{\pi^2}(w-\hat{m}^3)\,.
\end{equation}
Here we have used that the five-dimensional Newton constant is related to the number of D3-branes via $G_N = 4\pi /(N^2g^3)$.

%%%%%%%%%%%%%%%%%%%%%%
\subsection{The GPPZ solution}
\label{subsec:GPPZ}
%%%%%%%%%%%%%%%%%%%%%%

The GPPZ solution \cite{Girardello:1999bd} solves the BPS equations of the four-scalar model and has the explicit form\footnote{The map to the scalar fields used in \cite{Girardello:1999bd} is  $\mu=\e^{\sigma_{\rm GPPZ}}$ and $\nu=\e^{m_{\rm GPPZ}/\sqrt{3}}$.} \cite{Bobev:2018eer}
\begin{equation}\label{eq:GPPZ5d}
\begin{split}
\dd s_5^2 &= \frac{4}{g^2 t^2} \left(\dd t^2 + \left(1-t^2\right) \left( 1-\lambda^2 t^6 \right)^{1/3}\dd s_4^2 \right)\,,\\
z_1 &= \rmi \frac{\mu - \nu^3}{\mu + \nu^3},\quad z_2 = \rmi \frac{1- \mu \nu}{1+\mu \nu},\quad \mu = \sqrt{\frac{1+ \lambda t^3}{1-\lambda t^3}},\quad \nu = \sqrt{\frac{1+t}{1-t}}\,,
\end{split}
\end{equation}
where $t=\hat{m}\exp\left(-g r/2 \right)$ is a new radial variable and $\hat{m}$ is defined in \eqref{UVexpflat}. Note that the scalars $z_{1,2}$ are purely imaginary, so only two of the four real scalars have a non-trivial profile. The solution is asymptotic to the AdS$_5$ vacuum as $t\to 0$. Expanding near the AdS boundary we can relate the integration constant $\lambda$ to the UV parameters $\hat{m}$ and $w$ in \eqref{UVexpflat} 
\be
(\lambda+1)\hat{m}^3= w\,.
\ee
Combining this with \eqref{eq:gauginovevhol} we find that the vev of the operator dual to $\phi_4$ is $\langle {\cal O}_{\phi_4} \rangle = \frac{N^2}{\pi^2}\hat{m}^3\lambda$.

The metric in \eqref{eq:GPPZ5d} has a naked singularity at $t=1$ which corresponds to the IR regime of the dual gauge theory. The structure of the singularity depends on the value of $\lambda$. It was argued by Gubser that physically acceptable naked singularities in the context of holography should have an on-shell value of the scalar potential which is bounded from above \cite{Gubser:2000nd}. Applying the Gubser criterion to the solution in \eqref{eq:GPPZ5d} we find that the naked singularity is acceptable for $\left|\lambda\right|\leq 1$.\footnote{The GPPZ solution is invariant under the simultaneous action $\lambda\to-\lambda$ and $t\to -t$. Since the UV is always located at $t=0$ solutions can either have $t\ge0$ or $t\le0$, but not both. We can always choose the coordinate $t$ to be positive but then we must consider any value $-1\leq \lambda \leq 1$.} From now on we focus only on the physically acceptable values of $\lambda$ and we analyze the structure of the naked singularity in detail when we uplift the GPPZ solution to ten dimensions.

%%%%%%%%%%%%%%%%%%%%%%
\subsection{Looking for a chiral condensate}
\label{subsec:chiralcond}
%%%%%%%%%%%%%%%%%%%%%%

The BPS equations of the four-scalar model are compatible with a non-trivial vev for the operator dual to the scalar $\alpha$, i.e. a non-trivial chiral condensate in $\mathcal{N}=1^*$. It is thus natural to ask whether there are supersymmetric gravitational domain wall solutions which obey the Gubser criterion and have a non-trivial profile for the scalar $\alpha$. Unfortunately the general BPS equations for the four-scalar model in \eqref{5DBPSflat} do not admit analytic solutions and to answer this question we have to resort to perturbation theory and a numerical analysis. It is a daunting task to systematically explore the parameter space $(m,w,v,\varphi_0)$ as introduced in \eqref{UVexpflat} and construct numerical solutions for all values of the UV parameters. We circumvent this by taking a slightly different approach. All domain wall solutions for which at least one scalar flows have a naked singularity in the IR region. We are only interested in acceptable naked singularities as dictated by the Gubser criterion. We therefore start by classifying the possible singular behavior in the IR region for all solutions of the BPS equations and perform a series expansion of the BPS equations around these singular IR solutions. This analysis proves sufficient to understand whether a given naked singularity obeys the Gubser criterion without the need to fully integrate the BPS equations.

Performing this analysis, we conclude that all domain wall solutions in which $\alpha$ \emph{and} one of $\phi, \phi_4$ have non-trivial profiles are either unphysical due to the Gubser criterion, or cannot be connected to the UV AdS$_5$ region.  The inability to connect the latter flows to  AdS$_5$ is due to an intricate structure in the superpotential $\mathcal{W}$ when both $\alpha$ and one of $\phi$ or $\phi_4$ are non-vanishing. Effectively, the superpotential partitions the scalar domain in two regions, one that contains the AdS$_5$ vacuum solution and one that does not. The physically acceptable naked singularities with non-vanishing $\alpha$ flow into the region of the scalar manifold without the AdS$_5$ vacuum and terminate on a line where the superpotential vanishes. This is depicted in Figure~\ref{gubserplot} and some more details of our analysis can be found in Appendix \ref{5DBPSeq}. Therefore we see that the only physically acceptable domain wall solution (with flat slicing) of the four-scalar model with non-trivial profile for the scalars  $\phi$ or $\phi_4$ is the GPPZ solution, which has $\alpha=0$ and $\varphi=\text{const}$.

There is a regular, \emph{analytic} solution of the four-scalar model with a non-trivial profile for the scalar $\alpha$, but it has $\phi=\phi_4=0$, and corresponds to a particular RG flow on the Coulomb branch of $\mathcal{N}=4$ SYM \cite{Freedman:1999gk}. For completeness we present this solution in Appendix \ref{5DBPSeq}. It is important to note that the discussion above was restricted to holographic domain walls with flat slicing, i.e. the four-dimensional metric in \eqref{eq:5dmetansatz} is that on Minkowski space. There are smooth supersymmetric domain wall solutions of the four-scalar model with $S^4$ slicing constructed in \cite{Bobev:2016nua} which we discuss in some detail in Appendix~\ref{app:S4}.
\begin{figure}[h]
\centering
\begin{overpic}[scale=0.34]{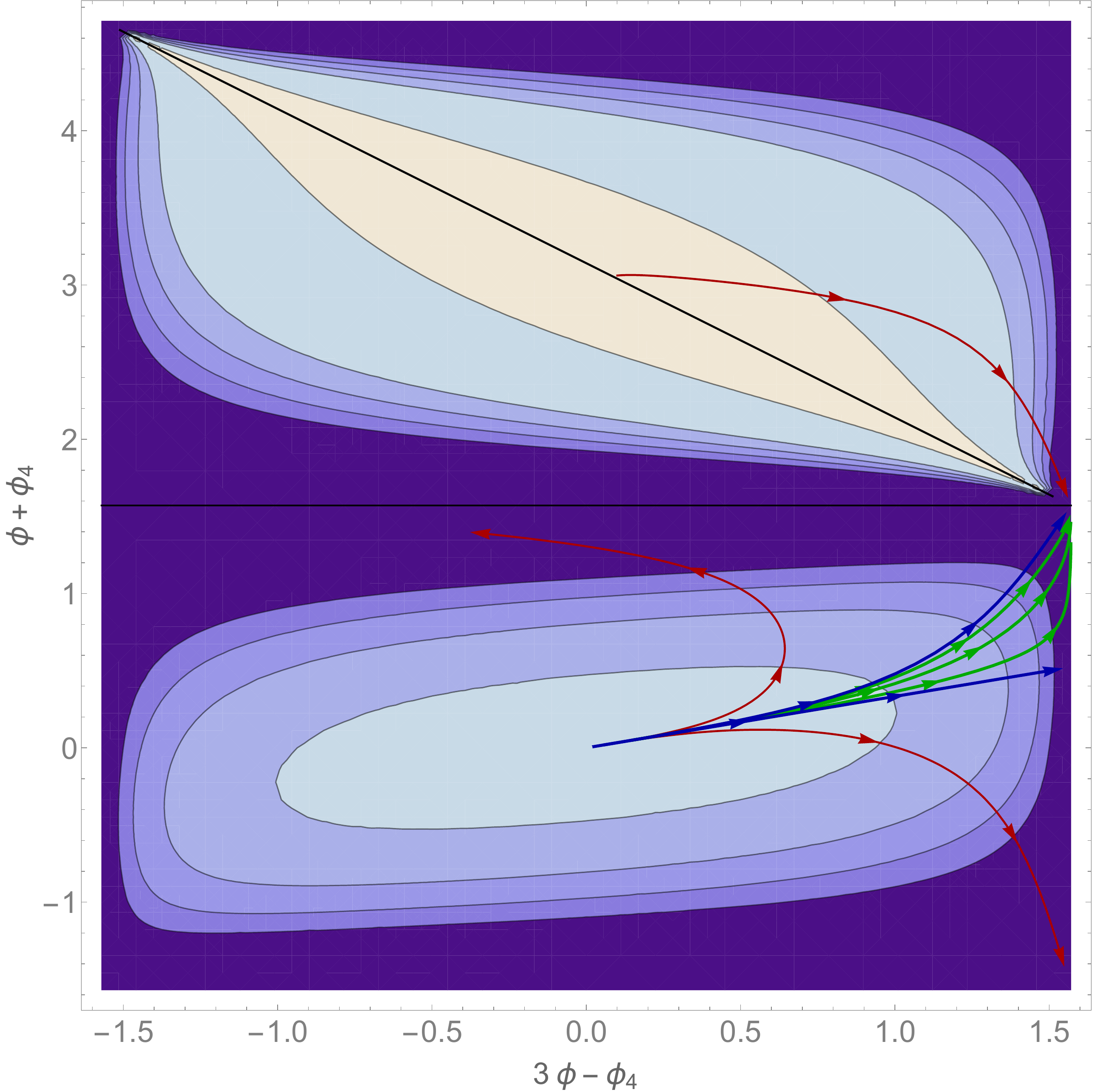}
\put(53,30.3){{\Large $\bullet$}}
\end{overpic}
\caption{Contour plot of $\e^{\cal K}{\cal W}\overline{\cal W}$ as a function of the scalars $\phi$ and $\phi_4$ on the surface $\alpha=0$. We restrict the plot to the fundamental domain of the two scalars. The AdS$_5$ solution has vanishing scalar fields and corresponds to the solid dot.  The coloured curves represent BPS domain wall flows with non-trivial scalars projected onto the $(\phi,\phi_4)$ plane. The two blue lines correspond to GPPZ flows with $\lambda=\pm1$, the GPPZ flows with $-1<\lambda<1$ all lie in between the two blue curves and are represented by the green curves. The two red lines denote unphysical domain wall flows according to the Gubser criterion which asymptote to the UV AdS$_5$ region. The red line in the upper half of the domain corresponds to an acceptable singular flow which does not connect to the AdS$_5$ solution. The Coulomb branch solution with $\phi=\phi_4=0$ is not visible on the plot since it starts from AdS$_5$ and extends in the direction orthogonal to the $(\phi,\phi_4)$ plane.}\label{gubserplot}
\end{figure}

%%%%%%%%%%%%%%%%
\section{The ten-dimensional solution}
\label{sec:10Dsolution}
%%%%%%%%%%%%%%%%

The five-dimensional GPPZ solution in \eqref{eq:GPPZ5d} can be uplifted to a solution of type IIB supergravity using the explicit uplift formulae in \cite{Baguet:2015sma}. This was done in \cite{Bobev:2018eer} and we summarize the relevant results below. In \cite{Petrini:2018pjk}, which appeared simultaneously with \cite{Bobev:2018eer}, a similar strategy was pursued and the full uplift of the GPPZ solution was also presented.\footnote{It should be noted that a partial uplift of the GPPZ solution was found in \cite{Pilch:2000fu} where the ten-dimensional metric and the axio-dilaton were written explicitly.}  He we briefly point out that in order to apply the uplift formulae of \cite{Baguet:2015sma}, one must make a choice of coordinates on the $S^5$, and there are many choices which are compatible with the $\SO(3)$ symmetry.  The uplifts given in \cite{Bobev:2018eer} and  \cite{Petrini:2018pjk} make different choices for these $S^5$ coordinates, and as a result some of the ten-dimensional fields of type IIB supergravity appear differently; however, we have checked explicitly that the two presentations of the uplift agree once one takes into account the difference in coordinates.  For completeness in Appendix~\ref{app:S5} we provide the explicit relation between the coordinates used in this paper and the ones in \cite{Pilch:2000fu} and \cite{Petrini:2018pjk}.

The solution in \cite{Bobev:2018eer} was written explicitly as a deformation of AdS$_5\times S^5$ and the coordinates on $S^5$ were chosen to reflect the $\SO(3)$ symmetry of the five-dimensional supergravity solution and the dual $\mathcal{N}=1^*$ gauge theory. The round metric on $S^5$ in these coordinates takes the form of a $\U(1)$ fibration over $\mathbf{C}  P^2$
\be\label{eq:metroundS5}
\dd\widehat{\Omega}_5^2 = \dd s_{\mathbf{C} P^2}^2 + (\dd\alpha+\sin2\chi~\sigma_3)^2\,,\qquad \dd s_{\mathbf{C}  P^2}^2 = \dd \chi^2 +
\sin^2\chi~\sigma_1^2+\cos^2\chi~\sigma_2^2+ \cos^22\chi~\sigma_3^2\,.
\ee
Here we have introduced the $\SO(3)$ left-invariant one-forms
\be
\begin{split}\label{eq:sigmaidef}
\sigma_1 =& -\sin \xi_2~ \dd\xi_1 + \sin\xi_1\cos\xi_2~\dd \xi_3\,,\\
\sigma_2 =& -\cos \xi_2~ \dd\xi_1 - \sin\xi_1\sin\xi_2~\dd \xi_3\,,\\
\sigma_3 =& -\dd\xi_2-\cos\xi_1~\dd\xi_3\,,
\end{split}
\ee
and the coordinates have the following ranges
\be
0\le\alpha\le 2\pi\,,\quad 0\le \chi\le \f{\pi}{4}\,,\quad 0\le \xi_1 \le \pi\,,\quad 0\le \xi_2 \le \pi\,,\quad 0 \le \xi_3 \le 2\pi\,.
\ee
In order to write down the type IIB supergravity solution in a relatively compact form we introduce the following functions
\begin{equation}\label{eq:Kdef}
\begin{split}
K_1 &= (1+t^2)(1-\lambda^2 t^8)+ 2t^2\left((1-\lambda^2 t^6)-\lambda t^2(1-t^2)\cos(4\alpha)\right)\cos2\chi\,,\\
K_2 &= (1+t^2)(1-\lambda^2 t^8)- 2t^2\left((1-\lambda^2 t^6)-\lambda t^2(1-t^2)\cos(4\alpha)\right)\cos2\chi\,,\\
K_3 &= 2\lambda t^4(1-t^2)\cos2\chi~\sin4\alpha\,,\\
K_4 &= (1+t^2)^2(1+\lambda t^4)^2  -4t^4(1+\lambda t^2)^2\cos^2 2\chi\,.
\end{split}
\end{equation}
The Einstein frame metric can then be written as
\begin{equation}\label{eq:10dmetexpl}
\dd s_{10}^2 = \frac{(K_1 K_2 - K_3^2)^{1/4}}{\sqrt{ g_s }}\left(\frac{\dd s_5^2}{\left(1-t^2\right) \left(1-\lambda^2 t^6\right)^{1/2}}+ \frac{4\left(1-\lambda^2 t^6\right)^{1/2}}{g^2(K_1 K_2 -K_3^2)}~\dd \Omega^2_5\right)\,,
\end{equation}
where $\dd s_5^2$ is the five-dimensional metric in \eqref{eq:GPPZ5d}. The squashed metric on $S^5$ can be written as
\begin{equation}\label{eq:sqspheremetexpl}
\begin{aligned}
\dd \Omega^2_5 =&~K_4 \dd\chi^2 - 4\lambda t^4(1-t^2)^2(\cos2\alpha~\dd\chi-\sin2\alpha~\cos2\chi~\sigma_3)^2\\
&- 4\lambda t^6~\dd (\cos2\alpha~\cos2\chi)^2  +\frac{(1-\lambda^2 t^8)^2(1-t^2)}{(1-\lambda^2t^6)}(\dd\alpha+\sin2\chi~\sigma_3)^2 \\
&+ \cos^22\chi(1+\lambda t^4)^2(4t^2 \dd \alpha^2+(1-t^2)^2\sigma_3^2)\\
&+(1-t^2)\big(\sin^2\chi~K_1\sigma_1^2 + \sin2\chi~ K_3\sigma_1\sigma_2 +\cos^2\chi~ K_2\sigma_2^2\big)\,.
\end{aligned}
\end{equation}
The axion and dilaton are given by
\begin{equation}\label{eq:axidilexpl}
\begin{split}
\rme^{\Phi} =& \frac{g_s(1+\lambda t^4)}{\sqrt{K_1 K_2-K_3^2}}\Big((1+t^2)(1-\lambda t^4)+2 t^2(1-\lambda t^2)\cos 2\chi~\cos 2\alpha\Big)\,,\\
C_0 =& -\frac{2 t^2 (1+\lambda t^2)(1-\lambda t^4)\cos 2\chi~\sin 2\alpha}{g_s(1+\lambda t^4)\big((1+t^2)(1-\lambda t^4)+2 t^2(1-\lambda t^2)\cos 2\chi~\cos 2\alpha\big)}\,.
\end{split}
\end{equation}
These can be combined into the complex axion-dilaton $\tau = C_0 + \rmi \rme^{-\Phi}$ which has nice transformation properties under the $\SL(2,\mathbf{R})$ symmetry group of type IIB supergravity. Note the appearance of the string coupling constant $g_s$ which is related to the coupling constant of the dual field theory via
\be\label{gYMdef}
g_\text{YM}^2 = 4\pi g_s\,.
\ee
The NS-NS and R-R two-forms can be written compactly as
\begin{equation}\label{eq:B2C2full}
\begin{split}
B_2 + \rmi g_s C_2 =& \frac{4}{g^2} \frac{t \e^{-\rmi \alpha}}{K_1 K_2 -K_3^2}\Big[\big( a_1 \dd \chi  + a_2 \sigma_3 -\rmi\left(1- \lambda^2 t^8 \right)\left( K_1 + K_2 \right) \sin 2\chi\, \dd \alpha \big) \w \Sigma\\
&- \big( a_3 \dd \chi+a_4 \sigma_3  - \rmi \left(1- \lambda^2 t^8 \right) \left( K_1 -K_2 - 2 \rmi K_3 \right)\sin 2\chi \, \dd \alpha \big)\w \overline{\Sigma}  \Big]\,,
\end{split}
\end{equation}
where we have defined the functions
\begin{equation}
\begin{aligned}
a_1=&- 2 \rmi K_3 \left(1+t^2\right)\left(1- \lambda^2 t^6 \right)\,,\\
a_2 =&\, \rmi \left(1+t^2\right) \big[(K_1-K_2) \left(1-\lambda ^2t^6\right)\cos 2 \chi-2 \left(1-\lambda ^2 t^8\right)^2\\
&-2t^2 \left(1+\lambda ^4
t^{12}-\lambda ^2t^4 \left(1+t^4\right) \right)\cos ^22 \chi  \big]\,,\\
a_3 =&\,4 t^4  \left(1-\lambda^2 t^4\right) \left(1-\lambda ^2 t^6-\lambda t^2 \left(1-t^2\right)  e^{4 \rmi \alpha }\right)\cos ^22 \chi \\
&- \left(1+t^2\right)^2 \left(1-\lambda ^2 t^8\right) \left(1-\lambda ^2 t^6+\lambda t^2  \left(1-t^2\right) e^{4 \rmi \alpha }\right)\,,\\
a_4 =&\, \rmi \left(1-t^2\right)^2  \left(1-\lambda ^2 t^8\right) \left(1-\lambda ^2 t^6-\lambda t^2 \left(1-t^2\right)  e^{4 \rmi \alpha }\right)\cos2 \chi \,, 
\end{aligned}
\end{equation}
and the complex one-form $\Sigma$ is given by
\begin{align}\label{eq:sigdef}
\Sigma = \rmi \sin \chi\, \sigma_1+ \cos \chi\, \sigma_2\,.
\end{align}
The R-R five-form is most compactly presented in terms of a four-form that only has legs along the Minkowski directions
\begin{equation}\label{eq:F5full}
F_5 =  -\f{1}{g^4g_s}(1+\star_{10})~\dd \left[\frac{\left(1-t^2\right)\left( 1-\lambda^2 t^8 \right)}{t^4\left(1-\lambda^2 t^6\right)^{1/3}}\, \dd x_0 \w \dd x_1 \w \dd x_2 \w \dd x_3\right]\,.
\end{equation}
The rank of the gauge group in the dual field theory, $N$, can be identified with the conserved D3-brane ``Page charge'' \cite{Marolf:2000cb}. This can be readily computed using the expressions above and one finds
\begin{equation}\label{Ndef} 
N = \f{1}{(2\pi\ell_s)^4}\int_{S^5} \left(F_5+\f12\left(C_2\w H_3 - B_2 \w \dd C_2\right)\right) = \frac{4}{g^4 g_s \ell_s^4 \pi}\,.
\end{equation}
We note also that the length scale of AdS$_5$ is given by $L=2/g$ and it is the same as the radius of the round $S^5$ in the UV. 

As expected from the five-dimensional GPPZ solution and from the dual $\mathcal{N}=1^{*}$ theory, the ten-dimensional background above preserves the $\SO(3)$ symmetry associated with the left-invariant forms $\sigma_i$. Furthermore we find that the solution is invariant under a discrete symmetry that involves both the $\SL(2,\mathbf{R})$ symmetry group of type IIB supergravity as well as a shift in the coordinate $\alpha$.\footnote{This symmetry is directly related to the discrete symmetry used to truncate the eight-scalar model in \cite{Pilch:2000fu} to the four-scalar model discussed in Section~\ref{sec:5Dsugra}} Specifically, we find that the S-duality transformation $\tau\to-1/(g_s^2\tau)$ combined with the shift $\alpha\to \alpha+\pi/2$ leaves all supergravity fields invariant. This invariance is clear for the metric and five-form since they are $\SL(2,\mathbf{R})$ singlets and only depend on  $\alpha$ through functions that are $\pi/2$ periodic. The two-forms are interchanged under S-duality 
\be\label{b2c}
B_2\to -g_s C_2\,,\qquad C_2\to g_s^{-1}B_2\,,
\ee
but combined with the shift of alpha, both forms are mapped to themselves. The same conclusion holds for the axion-dilaton $\tau$.  As explained in \cite{Bobev:2018eer} this  discrete symmetry gets enhanced to a $\U(1)$ symmetry for $\lambda=0$. Finally, we note that for $|\lambda| \to \infty$ the symmetry of the background above is enhanced to $\SU(3)\times \U(1)$, see \cite{Bobev:2018eer} for an explicit discussion. Since this value of $\lambda$ lies outside the range $|\lambda|<1$ allowed by the Gubser criterion we will not discuss it further.

%%%%%%%%%%%%%%%%
\subsection{The singularity for $|\lambda|<1$}
\label{subsec:lambdaless1}
%%%%%%%%%%%%%%%%

We have already noted that the family of five-dimensional solutions described in Section \ref{sec:5Dsugra} is singular as $t$ approaches $1$. The solutions are nevertheless physical when $|\lambda| \leq 1$ as we will argue. In this section we focus on the case where $\lambda$ is strictly smaller than one. As shown below, in this case the singularity can be attributed  to the presence of explicit smeared branes in the geometry. A similar conclusion was reached previously by Pilch and Warner in \cite{Pilch:2000fu}, however it was based on a partial uplift of the GPPZ solution in which only the metric and the axio-dilaton were given. We make this analysis more rigorous by studying the full set of type IIB supergravity fields. 

In the ten-dimensional solution the singularity as $t\to1$ is only present at a specific locus on the five-sphere, in particular for $|\lambda|<1$ the singularity is located  at the point $(t,\chi) = (1,0)$.\footnote{See Appendix~\ref{app:singularity} for a discussion of what precisely is meant by an expansion around this point.} The result is the following metric
\begin{equation}\label{Eq:zoomedmetric}
\begin{aligned}
\dd s_{10}^2 \approx& 
 \frac{8}{g^2 \sqrt{g_s}}H^{-1/4} \left[\left( 1-\lambda^2 \right)^{1/3} \dd s_{1,3}^2  +k(\alpha) \left(\dd \chi^2 + \chi^2 \dd\psi^2\right) \right]\\
 &\qquad\qquad + \frac{1}{2g^2 \sqrt{g_s}}H^{3/4} \left[\frac{4}{k(\alpha)} \left( \dd \alpha^2 +\dd \rho^2 +\rho^2\dd\Omega_2^2\right)\right]\,,
\end{aligned}
\end{equation}
where $\rho=1-t$, and thus $\rho\to0$, $\dd s_{1,3}^2$ is the metric on $\mathbf{R}^{1,3}$, and $\dd \Omega_2^2$ is the metric on the unit-radius round 2-sphere. We have also defined the functions 
\begin{equation}\label{eq:Hkdef}
H= \frac{2k(\alpha)}{\rho}\,, \qquad k(\alpha) \equiv \f{1-\lambda^2}{ 1+ 2\lambda \cos 4\alpha + \lambda^2 }\,.
\end{equation}
Notice that we have parametrized the $\SO(3)$ spanned by $\xi_{1,2,3}$ in \eqref{eq:sigmaidef} by the angle $\psi$ and the coordinates on $S^2$. The metric is singular along the entire circle parametrized by the $\alpha$ coordinate. The metric in \eqref{Eq:zoomedmetric} bears many similarities to the metric of a set of coincident five-branes in flat space\cite{Lu:1998vh}
\begin{equation}\label{eq:5branesmet}
\dd s_{10}^2 = h^{-1/4} \dd s_{6}^2 + h^{3/4} \dd s_{4}^2, \quad \text{where} \quad h = 1+  \frac{T}{r^2},
\end{equation}
where $\dd s_6^2$ denotes the brane world-volume, $\dd s_4^2$ denotes the space transverse to the branes and the coordinate $r$ denotes the distance from the stack of branes. In the harmonic function $h$ the parameter $T$ is related to the tension of the five-brane.

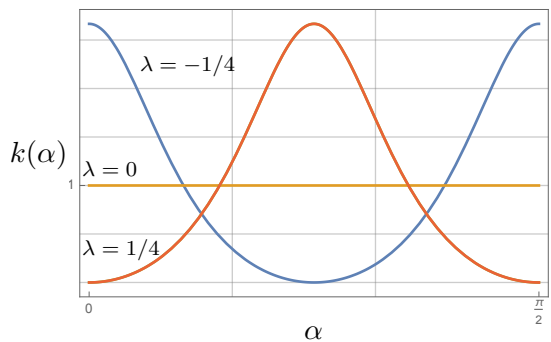
\begin{figure}[H]
\centering
\begin{overpic}[scale=0.5,unit=1bp,tics=20]{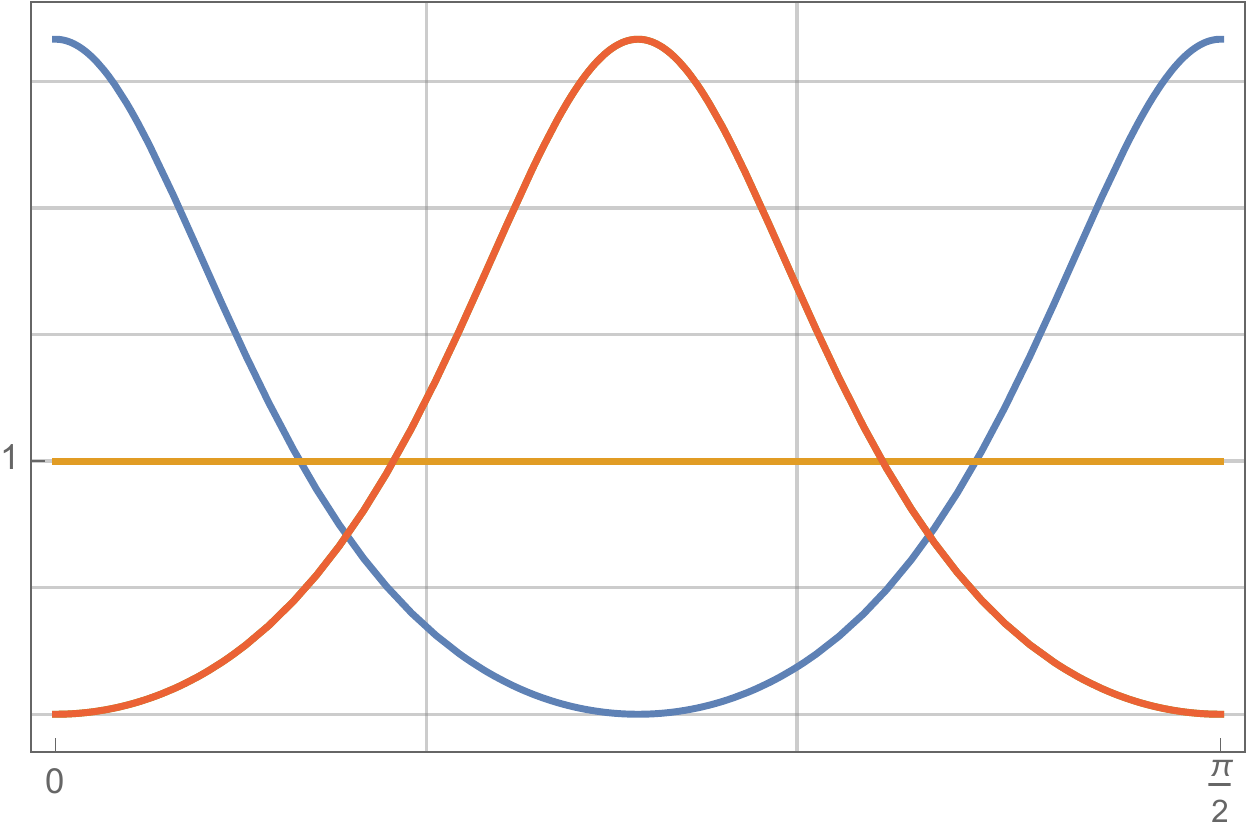}
\put(49,-2){$\alpha$}
\put(-12,35){$k(\alpha)$}
\put(15,54){\scriptsize$\lambda=-1/4$}
\put(3,32){\scriptsize$\lambda=0$}
\put(3,15.5){\scriptsize$\lambda=1/4$}
\end{overpic}
\caption{The function $k(\alpha)$ for three different values of $\lambda$. The function is $\pi/2$ periodic and should be extended to cover the entire range $0\le\alpha\le2\pi$. As $|\lambda|$ approaches 1 the function gets concentrated around the peaks at $\alpha=0+n \pi/2$ for $\lambda\to-1$ and $\alpha = \pi/4+n\pi/2$ for $\lambda\to1$, where $n\in\mathbf{Z}$.}\label{Fig:kfunction}
\end{figure}

The metric in \eqref{Eq:zoomedmetric} differs from the one in \eqref{eq:5branesmet} in several important ways. First, since we are already in the ``near-horizon'' limit we do not see the $1$ in the harmonic function as in \eqref{eq:5branesmet}. Second, we notice that the five-branes in \eqref{Eq:zoomedmetric} appear to be smeared since the degree of singularity of the harmonic function, $H$, is less than that of $h$. Indeed, the four-dimensional space transverse to the five-branes in \eqref{Eq:zoomedmetric} takes the form of a warped cylinder and at every point on the circle parametrized by $\alpha$ there sits a five-brane. The cylinder is warped by the $\pi/2$-periodic function $k(\alpha)$ in \eqref{eq:Hkdef}. This function also appears in as a prefactor in front of the the space spanned by $(\chi,\psi)$ in the six-dimensional world-volume of the five-branes. This space is a part of a two-dimensional compact submanifold of $S^5$ into which the five-branes polarize. The function $k(\alpha)$ therefore has a natural interpretation as the polarization radius of the five-branes. However its appearance in the harmonic function $H$ also suggests that it plays the role of the tension of the five-brane. The function $k(\alpha)$ plays an important role in the holographic interpretation of the geometry and we note here that the integral of it is independent of $\lambda$,
\be\label{eq:int}
\int k(\alpha)\dd\alpha = 2\pi\,.
\ee
In Figure~\ref{Fig:kfunction} we plot $k(\alpha)$ for various values of $\lambda$.
The rest of the type IIB supergravity fields are compatible with the interpretation of the singularity as a smeared stack of five-branes. The axion and dilaton take the form
\begin{align}
\rme^{\Phi} \approx g_s H^{1/2} \cos^2 \alpha, \quad C_0 \approx -\frac{\tan \alpha}{g_s}\,.
\end{align}
These can be combined into the complex type IIB axio-dilaton
\be
\tau \equiv  C_0 + \rmi\e^{-\Phi} \approx \frac{1}{g_s} \f{H^{1/2}\sin \alpha+\rmi \cos \alpha}{H^{1/2}\cos \alpha- \rmi \sin \alpha }\,.
\ee
We have chosen to write the expression for $\tau$ as a compact $\SL(2,{\bf R})$ rotation by an angle $\alpha$ of the value of $\tau$ at $\alpha=0$. This structure repeats itself for the two-forms
\be\label{eq:B2C2sing}
\begin{bmatrix} B_2\\  C_2\end{bmatrix} \approx \begin{bmatrix}
\cos \alpha & g_s\sin \alpha \\ 
-g_s^{-1}\sin \alpha & \cos \alpha
\end{bmatrix}\begin{bmatrix} 0\\ \tfrac{4}{g_sg^{2}}\vol_{S^2}\end{bmatrix}\,.
\ee
Finally the five-form flux takes the near-singularity form
\begin{equation}
\begin{aligned}
F_5 \approx \f{2(1-\lambda^2)^{2/3}}{g^4 g_s}(1+\star_{10}) \dd \rho \w \dd x_0 \w \dd x_1 \w \dd x_2 \w \dd x_3\,.
\end{aligned}
\end{equation}
The metric in \eqref{Eq:zoomedmetric} has the structure of a smeared distribution of five-branes along the coordinate $\alpha$. We can see this more explicitly by computing the five-brane charge density along the $\alpha$-circle. For the NS5-brane charge we integrate the near-singularity expression for $dB_{2}$ between two points on the circle and divide by the length of an infinitesimal circle-arc. The result is
\be\label{eq:palpha}
p(\alpha) = \f{1}{(2\pi\ell_s)^2} \lim_{\epsilon\to0}\f{1}{\epsilon}\int_{\bar\alpha=\alpha}^{\alpha+\epsilon}\dd B_2(\bar\alpha)  = \sqrt{\f{4 g_sN}{\pi}}\cos\alpha\,,
\ee
where we have written the final answer in terms of the field theory quantities using \eqref{gYMdef} and \eqref{Ndef}. A similar computation for the D5-brane charge leads to
\be\label{eq:qalpha}
q(\alpha) = - \sqrt{\f{4 N}{\pi g_s}}\sin\alpha\,.
\ee
We note that these charges are so-called ``Page'' charges and therefore should be quantized \cite{Marolf:2000cb}. This may appear puzzling since the angle $\alpha$ is a continuous angular coordinate. We believe that this puzzling behavior is an artefact of the large $N$ limit. It is natural to speculate that $1/N$ effects will lead to \emph{desmearing} of the five-branes and this will ultimately resolve the charge quantization puzzle. A similar effect in a different context was discussed in \cite{Tong:2002rq}. Finally we can also compute the D3-brane charge density along the $\alpha$ circle and we find that it is constant
\be
Q_\text{D3}(\alpha) = \f{N}{2\pi}\,.
\ee

To summarize, we have found that the naked singularity for $|\lambda|<1$ is due to a smeared line distribution of $(p,q)$-five-branes that carry D3-brane charge. Note that the total five-brane charge vanishes
\be
\int p(\alpha)\,\dd\alpha = \int q(\alpha) \,\dd\alpha=0\,,
\ee
This is fully compatible with the solution far away from the naked singularity which approaches AdS$_5\times S^5$ at asymptotic infinity. The magnitude of the localized five-brane charge is constant along the ring-like singularity the tension of the five-branes is not and is controlled by the function $k(\alpha)$ in \eqref{eq:Hkdef}. This is due to finite binding energy between the five-branes which is due to the non-trivial axion and dilaton to which the branes are sensitive \cite{Lu:1998vh,Witten:1995im}. The presence of the D3-brane charge near the singularity is entirely compatible with the dielectric brane effect of Myers \cite{Myers:1999ps}. The five-branes carry D3-brane charge since they are a result of the polarization of the D3-branes in the presence of the 2-form fluxes in \eqref{eq:B2C2sing}. 

%%%%%%%%%%%%%%%%
\subsection{Probe strings and line operators}
\label{Sec:LineOperators}
%%%%%%%%%%%%%%%%

To collect more evidence in favor of the above interpretation of the naked singularity in terms of polarized five-branes, we can study it with probe strings.  From the perspective of the dual gauge theory, these strings are dual to line operators, and play the role of the order parameters for deconfinement originally discussed in \cite{tHooft:1977nqb}, and re-emphasized for holographic field theories in \cite{Witten:1998zw}.  Thus a careful study of probe strings can reveal how to classify the field theory vacuum dual to our solution in terms of the vacua discussed in Section~\ref{Sec:FieldTheory}.  A similar approach was ventured in \cite{Girardello:1999bd} on their five-dimensional background, but as pointed out in \cite{Pilch:2000fu}, the five-dimensional approach may be misleading as it neglects the possibility for probe strings to couple to type IIB supergravity fluxes and/or develop a non-trivial profile in the $S^5$ directions.

We first review some standard facts about probe strings in a holographic context.  Vacuum expectation values of line operators can be computed in AdS/CFT by inserting probe strings into the geometry and computing their (regularized) on-shell action  \cite{Maldacena:1998im,Rey:1998ik,Drukker:1999zq}. These strings ``hang'' from the boundary into the bulk geometry. The boundary conditions of the string are determined by the type of line operator of interest. Probe strings in type IIB string theory come in two flavors, both of which play a role in our discussion. First, we have the fundamental string which is charged with respect to the Kalb-Ramond field $B_2$. Second, we have a D1-brane which is charged under $C_2$. A bound state of $m$ fundamental strings  and $n$ D1-branes is referred to as $(m,n)$-string and is charged with respect to a linear combination of $B_2$ and $C_2$. The fundamental string is dual to a Wilson line operator whereas the D-string is dual to a 't~Hooft line operator, both in the fundamental representation of the gauge group. An $(m,n)$-string is dual to a line operator which can be thought of as a product of Wilson and 't~Hooft line operators. The probe string has a certain position on the five-sphere and so the dual line operator transforms non-trivially under the $\SO(6)$ R-symmetry of the UV $\mathcal{N}=4$ SYM theory.  As pointed out in \cite{Maldacena:1998im}, this coupling to the R-symmetry arises because line operators in $\mathcal{N}=4$ SYM involve, in addition to the usual gauge field holonomy, a second term built from the scalar fields. For example, a Wilson line is given by
\begin{equation}\label{Eq:wilsonlinedef}
W\left[C,\theta\right] = \text{Tr} \, \text{Pexp} \int_C \rmi \left( A - \theta^I X_I \right)\dd s,
\end{equation}
where $C$ specifies the contour of integration and $\theta^I$ are six additional functions which describe the the path of this contour through R-symmetry space (effectively, on the internal $S^5$).  A similar coupling to the scalars $X_I$ appears in the 't~Hooft line operators. For particular choices of $\theta^I$ and $C$, these operators may preserve a subset of the supercharges in ${\cal N}=4$ SYM theory, see for example \cite{Zarembo:2002an}. However, for the ${\cal N}=1^*$ SYM theory of interest here, all line operators break supersymmetry.

\begin{figure}
\centering
\tikzset{
	particle/.style={thin,draw=black, postaction={decorate},
		decoration={markings,mark=at position .55 with {\arrow[black,scale=2]{stealth}}}},
	length/.style={thin,draw=black, postaction={decorate},
		decoration={markings,mark=at position 1 with {\arrow[black,scale=2]{stealth}}}}
}

\begin{tikzpicture}[node distance=1cm and 1.5cm]

\coordinate[label=] (e1);
\coordinate[below=4cm of e1] (e2);
\coordinate[right=of e1] (e3);
\coordinate[below=4cm of e3] (e4);

\coordinate[right=0.1cm of e1](aux1);
\coordinate[right=1.4cm of e1](aux2);

\coordinate[left=2cm of e1](aux3);
\coordinate[below=4cm of aux3](aux4);

\draw[particle] (e1) -- node[label=left:$\bar q$] {}(e2);
\draw[particle] (e4) -- node[label=right:$q$] {}(e3);

\draw[length] (aux2) -- node[label=above:$L_{q\bar q}$] {}(aux1);
\draw[length] (aux1) -- (aux2);

\draw[length] (aux4) -- node[label=left:$x_0$] {}(aux3);
\end{tikzpicture}
\caption{A quark-antiquark pair at a distance $L_{q\bar{q}}$ from each other.}\label{Fig: q qbar pair}
\end{figure}
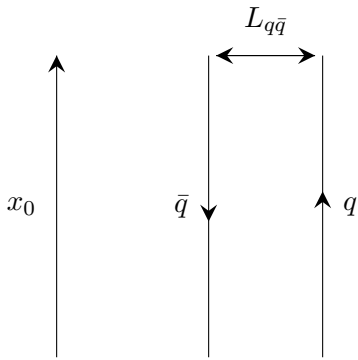

The vev of the line operator in \eqref{Eq:wilsonlinedef} encodes information about the vacuum structure of the gauge theory. This is somewhat analogous to the way in which the quark-anti-quark potential is sensitive to confinement. In ${\cal N}=1^*$ SYM there are no matter fields in the fundamental representation, but one can mimic the notion of ``quark-anti-quark potential'' by studying a rectangular loop operator. In particular, we choose a closed rectangular contour that extends along the time direction, $x_0$, with length $L_0$ and along one of the spatial directions with length $L_{q\bar q}$. We take $L_0\gg L_{q\bar q}$ such that the line operator resembles two disconnected line operators associated to a ``quark'' and an ``anti-quark'' with separation $L_{q\bar q}$, see Figure~\ref{Fig: q qbar pair} for an illustration. We emphasize that this fictitious ``quark-anti-quark'' pair is only a tool to visualize our setup. In the limit $L_0\gg L_{q\bar q}$ the vev of the Wilson line takes the form
\begin{equation}\label{Eq: Wilsonline vev and quark potential}
\left\langle W\left[C\right] \right\rangle \propto \rme^{-V_{q\bar q}\left(L_{q\bar q}\right) L_0}\,,
\end{equation}
where $V_{q\bar q}$ can be thought of as the \emph{quark-antiquark potential}. The behavior of this potential for sufficiently large $L_{q\bar q}$ encodes properties of the gauge theory vacuum. If the potential grows linearly, $V_{q\bar q} \sim L_{q\bar q}$, the Wilson loop vev displays an area law which indicates confinement. If the potential approaches a constant, $V_{q\bar q} \sim \text{const}$, the quarks are  screened.

Our goal is to compute the potential $V_{q\bar q}$ using probe strings in the explicit solution \eqref{eq:Kdef}-\eqref{eq:F5full}. This is done by fixing the contour of the probe string on the boundary of AdS$_5$ as in Figure~\ref{Fig: q qbar pair} and finding a configuration in the bulk which minimizes the string action. The potential $V_{q\bar q}$ is then extracted from the regularized on-shell action of this probe string. We emphasize that in the calculation below we use the full type IIB supergravity solution in \eqref{eq:Kdef}-\eqref{eq:F5full} and not the near-singularity background discussed in Section~\ref{subsec:lambdaless1}.

The action for a probe $(m,n)$-string takes the form
\begin{equation}\label{Eq: NG action}
S_{(m,n)} =-\frac{1}{2\pi\ell_s^2} \int \left[\dd^2 \sigma \sqrt{(n^2\e^{-\Phi}+\e^{\Phi}(m-nC_0)^2) | P\left[g_{MN}\right]|}  - P\left[m B_2+nC_2\right]\right]\,,
\end{equation}
where $P[\cdots]$ denotes the pullback of the ten-dimensional field onto the string world-volume. Notice that the tension of the $(m,n)$-string is not just the sum of the tensions of $m$ fundamental strings and $n$ D1-strings. This is similar to the case of $(p,q)$ five-branes discussed above where the binding energy contributes non-trivially.

Motivated by the discussion above we embed the string worldsheet in the ten-dimensional geometry by identifying the world-sheet time coordinate $\sigma_0$ with the four-dimensional boundary time coordinates $x_0$ and assume that the embedding of the string does not depend on the time-coordinate. This implements the static configuration of the ``quarks'' discussed above and leads to a dramatic simplification. Since, the 2-forms in \eqref{eq:B2C2full} have no legs along the four-dimensional space-time on the boundary, their pull-back necessarily vanishes. We are therefore left to compute the determinant of the pullback of the metric which reads
\begin{equation}\label{Eq: pull back on worldsheet}
\frac{16\left(K_1 K_2 - K_3^2\right)^{1/2}}{g_sg^4T^4\left(1-\lambda^2 T^6\right)^{1/3}}\Big( \dot{X}^2 + \frac{\dot{T}^2}{\left(1-T^2\right)\left( 1-\lambda^2 T^6 \right)^{2/3}}+ \frac{T^2 \sqrt{1-\lambda^2 T^6}}{K_1 K_2 -K_3^2} G_{mn} \dot{\Theta}^m \dot{\Theta}^n\Big)\,.
\end{equation}
Here $T\left(\sigma\right)$ and $\Theta^{m}\left(\sigma\right)$ with $m=1,\ldots,5$ are functions of the spatial world-sheet coordinate, $\sigma=\sigma_1$, which encode how the string is embedded along the radial coordinate of AdS$_5$ and the five angles of the $S^5$. We have fixed the embedding of the string along two of the spatial directions along the AdS$_5$ boundary by setting $x_2=x_3=0$. Finally, the function $X\left(\sigma\right)$ encodes the string embedding along the $x_1$ direction in the AdS$_5$ boundary. We use a dot to denote the derivative with respect to $\sigma$.

Since the metric $G_{mn}$ in \eqref{Eq: pull back on worldsheet} is positive definite we can conclude that one way of extremizing the string action is to take the angles $\Theta^m$ to be constants as a function of $\sigma$. Note however that \emph{extremizing} the string action does not guarantee that the resulting solution provides the minimum on-shell action. Indeed, we show below that strings that have non-constant angles on the $\Theta^m$  can sometimes have lower energy than those with constant angles. For $(m,n)$-strings with constant angles $\Theta^m$ it is simple to minimize the action and find the following four distinct solutions
\begin{equation}\label{twostringsols}
\chi=0~,\quad \alpha \in \{0,1,2,3\} ~\pi/4~,\quad \text{and}\quad n = m g_s \sin^2(2\alpha)\,.
\end{equation}
Here we have used the discrete symmetry discussed above \eqref{b2c} to relate all other solutions to these four. Note that this symmetry acts non-trivially on the probe string itself since it involves an $S$-duality transformation in type IIB string theory. For $\lambda=0$ we naively find that $\alpha$ is completely unrestricted and there appear to be more solutions. However, since exactly for $\lambda=0$ the discrete symmetry is enhanced to a continuous $\U(1)$ symmetry, all the solutions are in fact equivalent to those in \eqref{twostringsols}. 
For all value of the angles in \eqref{twostringsols} the dimensionless string action takes the form
\be\label{mnactionsimple}
s\equiv -\f{2\pi \ell_s^2 g^2g_s}{L_0\sqrt{m^2g_s^2 + n^2}}S_{(m,n)} =\int \dd \sigma~ \zeta(T)\, \left[\dot{X}^2 + \frac{4 \dot{T}^2}{\left(1-T^2\right)\left(1-\lambda^2 T^6\right)^{1/3}}\right]^{1/2}~,\\
\ee
where $\zeta(T)$ is a non-trivial function of the scalar $T$ which takes a different form depending on the choice of angle in \eqref{twostringsols}. Since the action is entirely independent of time, we have performed the integral over the $x_0$ direction resulting in the explicit factor of $L_0$.\footnote{It is perhaps instructive to think of $s$ as an action density.} In the subsequent discussion we focus on the two solutions, $\alpha=0,\pi/2$ corresponding to fundamental, i.e. $(1,0)$, strings sitting at the two different positions on the five-sphere. These two solutions show qualitatively different behavior. The other two solutions, $\alpha = \pi/4,3\pi/4$, correspond to $(1,1)$ strings and display similar behavior to the first two. We can parametrize the two fundamental string solutions in terms of the constant value of $\alpha$ which leads to the following expression for the function $\zeta$
\be
\zeta^2 = \f{(1+\lambda T^4)\big((1+T^2)(1-\lambda T^4)+2T^2(1-\lambda T^2)\cos2\alpha\big)}{T^4(1-\lambda^2T^6)^{1/3}}\,.
\ee
We have arrived at a simple classical mechanics problem in one dimension with two variables $X(\sigma)$ and $T(\sigma)$.\footnote{The role of time is played by the spatial coordinate $\sigma$ on the string world sheet.} Let ${\cal L}$ be the Lagrangian of this one-dimensional problem, given by $s=\int {\cal L}\,\dd \sigma$ in \eqref{mnactionsimple}. The momenta conjugate to the variables $X$ and $T$ are 
\begin{equation}
p = \zeta^2 \frac{\dot X}{\mathcal L}\,,\qquad P_T =  \frac{4\zeta^2}{\left(1-T^2\right) \left( 1- \lambda^2 T^6 \right)^{1/3}}\frac{\dot T}{\mathcal L}\,.
\end{equation}
Note that since the Lagrangian is independent of $X$ its associated momentum $p$ is conserved and can be used to label the solutions. Furthermore, due to the reparametrization invariance of the Lagrangian, the Hamiltonian vanishes 
\begin{equation}
\mathcal H = p \dot{X} + P_T \dot{T} - \mathcal L= 0\,.
\end{equation}
It proves useful to parametrize the solutions of this one-dimensional problem in terms of the action integral itself, i.e. use a ``proper time'' parametrization such that the action is $\dd s = \mathcal L \,\dd \sigma$. This reduces the Hamiltonian constraint to a simple effective potential problem with zero total energy
\begin{equation}\label{Eq:classicalproblem}
 \f12 \left(\f{\dd T}{\dd s}\right)^2 +  V_\text{eff}= 0\,,\quad \text{with}\quad  V_\text{eff}  = \f{(1-T^2) (1-\lambda^2 T^6)^{1/3} ( p^2 -\zeta^2)}{8\zeta^4}\,.
\end{equation}
For each value of the parameter $p$ we want to find a solution to the classical mechanics problem. In particular we are interested in solutions which have a turning point where the potential energy vanishes and the velocity can switch sign. Such solutions describe a string profile with the two ends of the string ``anchored'' to the AdS$_5$ boundary which extends into the bulk. The turning point is found for some $T=t_0$ when $p^2 = \zeta^2(t_0)$ and its location represents how deep in the bulk the string extends. It is more convenient to label the solutions not by the conserved momentum $p$ but rather by the coordinate of the turning point $t_0$.

The quantities of physical relevance for our purposes are the renormalized on-shell action in \eqref{mnactionsimple} and the boundary separation between the two ``quarks'' $L_{q\bar q}$. Using \eqref{Eq:classicalproblem} one finds the following expressions for these quantities
\be\label{eq:sLren}
s^\text{ren}(t_0)=\lim_{\epsilon\to 0}\left[\int_\epsilon^{t_0}\f{\dd T}{\sqrt{-2V_\text{eff}}} - \f{2}{\epsilon}\right]~,\quad L_{q\bar q}(t_0) = 2\zeta(t_0)\int_0^{t_0}\f{\dd T}{\zeta^2\sqrt{-2V_\text{eff}}}\,.
\ee
Notice that we have multiplied these expressions by 2 since the full world-sheet is symmetric around the turning point $t_0$. Note also that the naive on-shell action in \eqref{mnactionsimple} diverges near the AdS$_5$ boundary, $t=\epsilon\to0$. To remedy this we included in \eqref{eq:sLren} the standard holographic counterterm to regularize the on-shell string action \cite{Maldacena:1998im,Drukker:1999zq,Rey:1998ik}. This counterterm ensures that we obtain a finite expression for the on-shell action as we take the limit $\epsilon\to0$. The integral for $L_{q\bar q}$ does not require regularization. The integrals in \eqref{eq:sLren} can be performed numerically and we discuss the results below.

\begin{figure}[ht]
\centering
\includegraphics[scale=0.75]{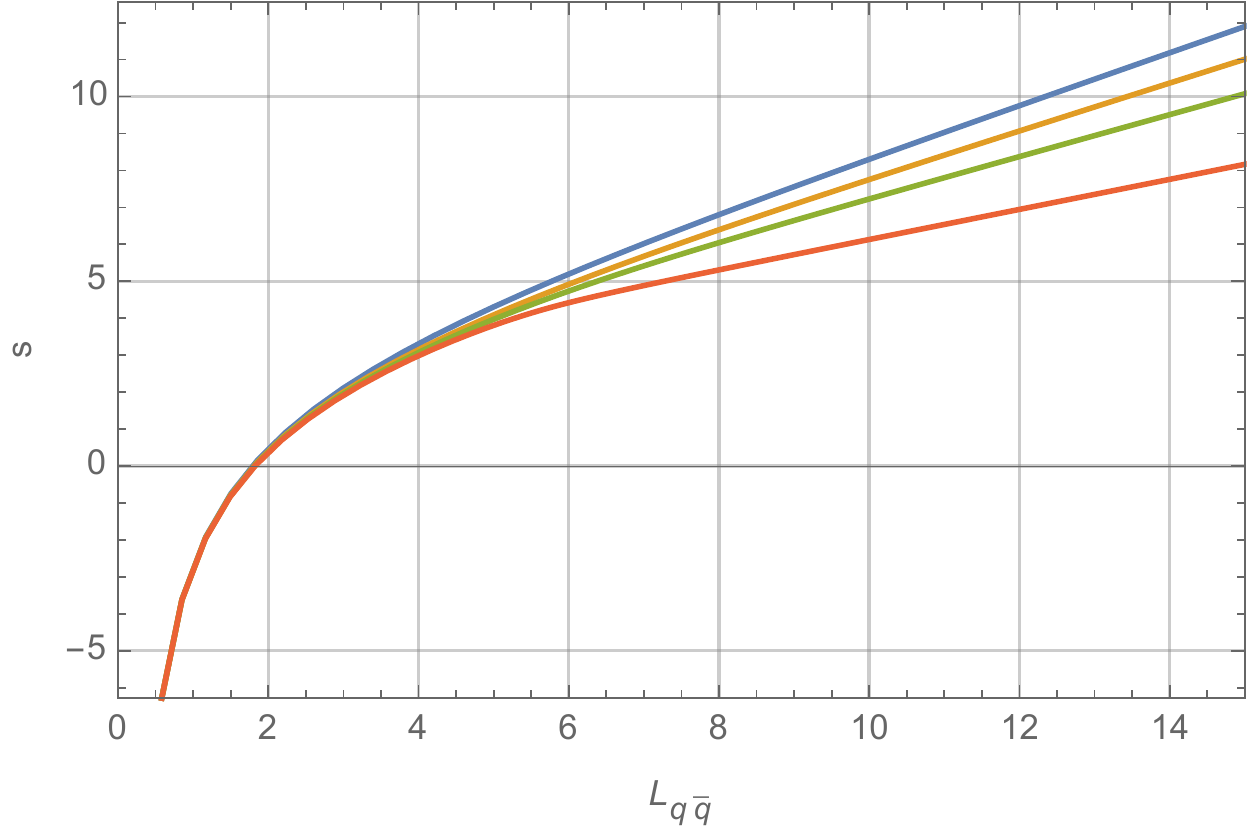}
\caption{The renormalized on-shell action for the fundamental string at $\alpha=0$, as a function of  $L_{q\bar q}$. The different lines correspond to different values of $\lambda$, namely $\lambda = (0,0.5,0.7,0.9)$, from top to bottom.}\label{Fig:SofLF1}
\end{figure}

As we emphasized above the calculation for the four different solutions in \eqref{twostringsols} can be treated simultaneously, however it turns out that the results are qualitatively different and thus we discuss them separately. For a fundamental string, i.e. $(m,n)=(1,0)$, at $\alpha=0$ we find that for large enough separation length of the quark-anti-quark pair the on-shell action grows linearly, as is shown in Figure~\ref{Fig:SofLF1}. 
The linear behavior in Figure~\ref{Fig:SofLF1} can be understood analytically by studying a string worldsheet formed by three straight lines.\footnote{We emphasize that this string profile solves the equations of motion but is never energetically favored. We use it here only as an approximation which captures the linear behavior in Figure~\ref{Fig:SofLF1}.} A straight line that extends from the UV AdS$_5$ region to the singularity at $t=1$. This is described by \eqref{Eq:classicalproblem} with $p=0$. This worldsheet contributes nothing to $L_{q\bar q}$ in \eqref{eq:sLren} but it contributes a fixed $\lambda$-dependent value to the on-shell action in \eqref{eq:sLren}. The second straight line segment is the string worldsheet that stretches along the singularity at $t=1$. This leads to $T=1$ and thus $\zeta=4(1-\lambda^2)^{2/3}$. Finally, the third piece of the worldsheet is a copy of the first one. We therefore find the following expression for the on-shell action of this three-piece string as a function of $L_{q\bar q}$
\be\label{analytic}
s =  (1-\lambda^2)^{1/3}L_{q\bar q}+\lim_{\epsilon\to 0}\left(\int^{1}_{\epsilon} \f{2\zeta\dd T}{\sqrt{(1-T^2)(1-\lambda^2T^6)^{1/3}}}-\f{2}{\epsilon}\right)\,.
\ee
The coefficient of $L_{q\bar q}$ in \eqref{analytic} provides an excellent fit to the slope of the linear regime of the numerical on-shell action in Figure~\ref{Fig:SofLF1}. This slope is simply given by the tension of a straight fundamental string that forms a bound state with the polarized NS5-branes sitting at $t=1$. Using the field theory expectations discussed around \eqref{Eq: Wilsonline vev and quark potential} it might be tempting to interpret this linear behavior as a sign of a confining vacuum with the tension of the flux-tube given by the tension of fundamental string in the presence of the polarized NS5-brane. However, as we discuss below this interpretation is problematic.

\begin{figure}[ht]
\centering
\includegraphics[scale=0.75]{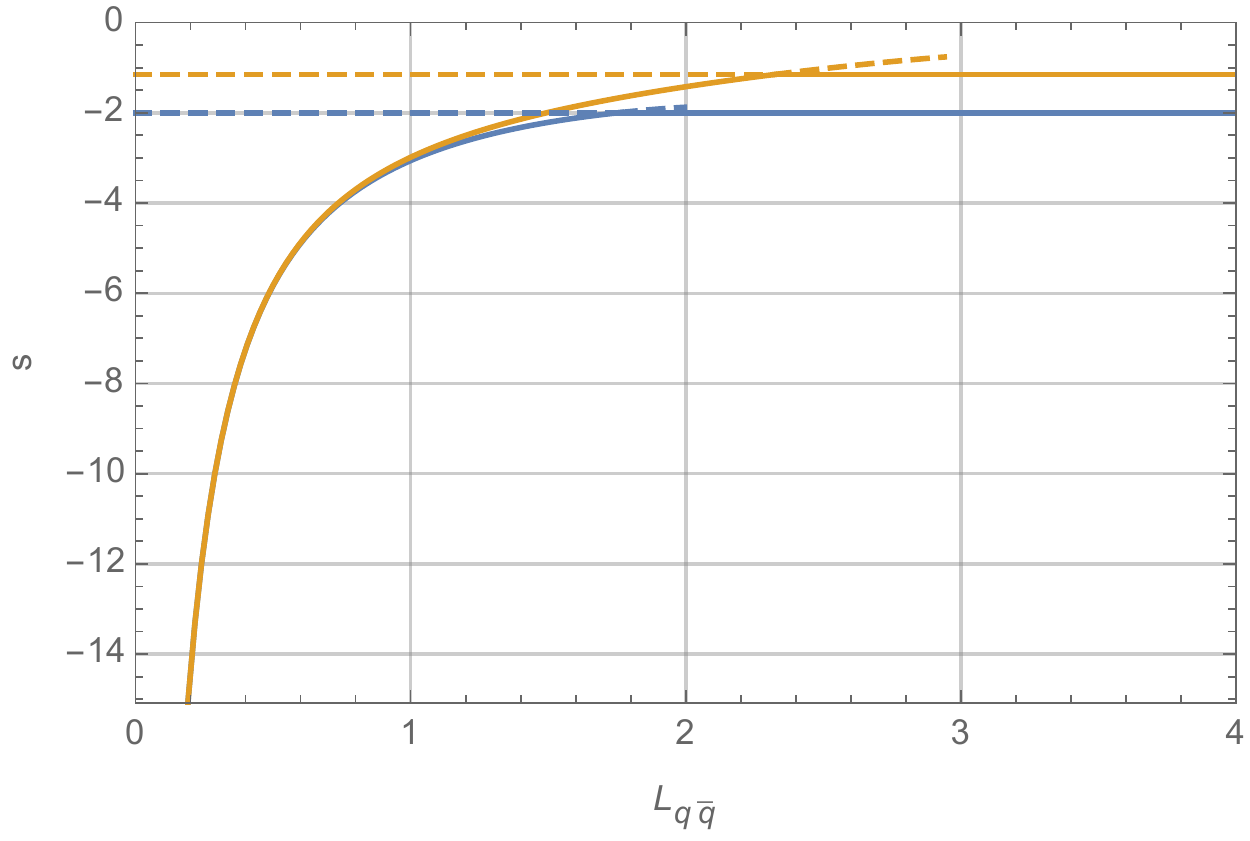}
\caption{The renormalized on-shell action for the fundamental string at $\alpha=\pi/2$ as a function of  $L_{q\bar q}$. The different lines correspond to $\lambda=(0,0.9)$ from bottom to up. For both values of $\lambda$ we show the two solutions discussed in the text above, the dominant and subdominant solutions are plotted with a solid and a dashed line, respectively.}\label{Fig:D1}
\end{figure}

\begin{figure}[ht]
\centering
\includegraphics[scale=0.06]{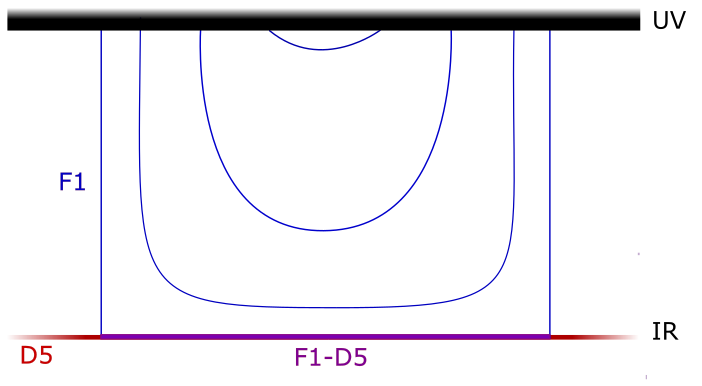}
\caption{A sample of $(1,0)$ string profiles with different values of $L_{q\bar{q}}$ at $\alpha=\pi/2$.}\label{Fig:F1D5}
\end{figure}

Now let us consider the other class of string solutions with $\alpha=\pi/2$. We again have a fundamental string, but according to \eqref{eq:palpha}-\eqref{eq:qalpha}, it is located at a value of $\alpha$ associated with polarized D5-branes. This changes the nature of the available string solutions. Similarly to the string at $\alpha=0$ we have a numerical solution with a turning point at $t_0<1$ for which one can compute the integrals in \eqref{eq:sLren}. However there is also another class of solutions composed of three straight segments just like the one described above \eqref{analytic}, see Figure~\ref{Fig:F1D5}. Contrary to the case with $\alpha=0$ this three-segment string may become energetically favored for some value of $L_{q\bar q}$. To decide which one of the two available solutions is dominant one has to compute the on-shell action of the string for each of them and choose the one with a lower value of the action. The on-shell action for the three-segment string solution can be computed analytically and is given by
\be\label{eq:sqqbar}
s^\text{ren}=-2(1-\lambda^2)^{2/3}+\f{2\lambda}{3} {}_2F_1\left(\f13,\f12,\f32,\lambda^2\right)-\f{6\lambda^2}{5} {}_2F_1\left(\f13,\f56,\f{11}6,\lambda^2\right)~.
\ee
Notice that this on-shell action is independent of $L_{q\bar q}$. This is because the tension of the bound state between the fundamental string and the polarized D5-brane vanishes and only the  two straight string segments connecting the UV AdS$_5$ to the naked singularity in the IR contribute to the action. For small values of $L_{q\bar q}$ the three-segment string solution is subdominant with respect to the numerical solution with a $t_0<1$ turning point. There is however a critical value of $L_{q\bar q}$ beyond which the three-segment string solution becomes dominant. This behavior is illustrated  in Figure~\ref{Fig:D1}. Comparing this behavior to the discussion around \eqref{Eq: Wilsonline vev and quark potential} we can conclude that the dual gauge theory is in a vacuum which exhibits screening. This leads to the interpretation that the on-shell action for the three-segment string in \eqref{eq:sqqbar} is equal to the ``quark-anti-quark'' binding energy in the dual gauge theory.

The analysis of the $\alpha=0$ and $\alpha=\pi/2$ solutions above leads to seemingly contradicting conclusions about the nature of the vacuum in the dual gauge theory, i.e. the $\alpha=0$ solutions indicate confining while the strings with $\alpha=\pi/2$ lead to a screening behavior. To clarify this note that for the fundamental string at $\alpha=0$ the energy grows linearly with $L_{q\bar q}$ whereas for $\alpha=\pi/2$ the energy reaches a maximum and stays constant no matter how much we increase $L_{q\bar q}$. This suggests that for large enough $L_{q\bar q}$ it is energetically favorable for the fundamental string placed at $\alpha=0$ to develop a profile along the $\alpha$ coordinate as it drops into the bulk such that near the singularity at $t=1$ one has $\alpha=\pi/2$. Indeed we have constructed examples of such solutions numerically but it is challenging to find a complete classification since one has to solve partial differential equations. Given the existence of these more general string configurations it is natural to expect that for any value of $\alpha$ near the AdS$_5$ boundary and for large enough quark separation, $L_{q\bar q}$, the dominant string solution will have $\alpha$ varying as a function of $t$ such that near the singularity at $t=1$ one finds $\alpha=\pi/2$. This then leads to the fundamental string binding with the polarized D5-brane at $\alpha=\pi/2$ exhibiting the screening behavior illustrated in Figure~\ref{Fig:D1}. This behavior is not restricted to fundamental strings but rather holds for all probe $(m,n)$-strings. For large enough separation, the $(m,n)$-string has a profile along the $\alpha$-angle such that for $t=1$ the value of the angle is $\tan\alpha=-g_s m/n$. This then leads to a bound state with a polarized $(n,-m)$ five-brane and a vanishing effective string tension. In the dual gauge theory this amounts to a screening behavior in the vacuum. 
%

%%%%%%%%%%%%%%%%
\subsection{The singularity for  $|\lambda|=1$}
\label{subsec:lambdaequal1}
%%%%%%%%%%%%%%%%

The type IIB supergravity solution has a naked singularity at $t=1$ which is physically acceptable, according to the criteria in \cite{Gubser:2000nd} and \cite{Maldacena:2000mw} for all values in the range $-1\le \lambda \le 1$. We have argued above how this singularity can be interpreted in terms of explicit polarized 5-brane sources when $|\lambda|<1$. The solutions with $|\lambda|=1$, however, need a separate treatment which we present here.

Fixing $\lambda=1$ and analyzing the background in \eqref{eq:Kdef}-\eqref{eq:F5full} one finds a naked singularity at $t \to 1$.\footnote{One can treat the $\lambda=-1$ in a very similar way.} To be more explicit it is convenient to define
\begin{equation}\label{eq:VWdef}
\begin{aligned}
w_1 =& \cos 2\chi \cos 2\alpha,\qquad w_2 = \cos 2\chi \sin 2\alpha\,,\\
V =& -4\left( w_1^2 + 4\left( w_2^2-1 \right) \right)\,, \qquad  W = - \frac{2\left(w_1^2 + 2w_2^2\right)}{\sqrt{w_1^2 + w_2^2}} \,.
\end{aligned}
\end{equation}
The ten-dimensional metric in the limit $t \to 1$ then takes the form
\begin{equation}\label{eq:10dmetl1}
\begin{aligned}
ds_{10}^2 &\approx \frac{V^{1/4}}{g^2\sqrt{g_s}\sqrt{6}(1-t)}\Bigg[ \left(10-6t\right)\dd t^2 + 48^{1/3}\left(1-t\right)^{4/3}\dd s_4^2 + \frac{6 (1+t)}{V}\dd w_2^2\\
&+ \frac{24\left(1-t\right)^2}{V}\left( \left(4 - W \right)\sin^2\chi\, \sigma_1^2 + \left( 4 + W \right)\cos^2 \chi\, \sigma_2^2 + \frac12\sin 4\chi \sin 4\alpha\, \sigma_1 \sigma_2 \right)\\
&+ \frac{ 16\left(1-t\right)^2}{V}\left(3 w_1 \left( 2\sin 2\alpha\, \dd \chi + w_1 \sigma_3 \right) + 4 \sin 2\chi \left( 2 \dd \alpha + \sin 2\chi \sigma_3 \right) \right)\sigma_3 \Bigg]\,.
\end{aligned}
\end{equation}
This form of the metric already suggests that any interpretation of the singularity as sourced by branes is difficult. The reason is that the whole five-sphere is singular for $t\rightarrow 1$. We are not aware of any brane distribution compatible with the symmetries in the problem that may lead to such a drastic singularity. 

One may be worried that this conclusion is due to an inappropriate choice of coordinates. To this end it is also useful to study the behavior of the background fluxes. The dilaton and axion near the singularity are given by
\begin{equation}
\begin{aligned}
\rme^{\Phi} \approx& \frac{2 g_s \left( 2 + \cos 2\chi \cos 2\alpha \right)}{\sqrt{V}}\,,\qquad C_0 \approx - \frac{2 \cos 2\chi \sin 2\alpha}{g_s \left(2 + \cos 2\chi \cos 2\alpha\right)}\,.
\end{aligned}
\end{equation}
It is clear that the axion and dilaton are regular for all points on the five-sphere except at 
\begin{equation}\label{eq:l1sing}
\chi=0\,, \qquad\qquad \alpha = \pi/4 + n \pi/2\,, \quad {\rm for} \quad n\in \mathbf{Z}\,.
\end{equation}
This is incompatible with any brane interpretation except  for a possible D1-brane located at the locus in \eqref{eq:l1sing}. However, D1-branes also source the R-R two form which can be read off from the following expression in the $t\to 1$ limit:
\begin{equation}
\begin{aligned}
&B_2 + \rmi g_s C_2 \approx\frac{4 \rme^{-\rmi \alpha}}{g^2 V}\Bigg[4\rmi \sin2\chi \left( \Big(4-3\cos2\chi+ \frac{(w_1+\rmi w_2)^2}{\cos2\chi}\Big) \overline\Sigma +8\rmi \sin\chi \sigma_1\right)\wedge \dd \alpha\\
& +\rmi V \Sigma \w \sigma_3 +\left(12\rmi w_1 \sin 2\alpha \,\Sigma-\left( 4 \rme^{4\rmi \alpha}+ 2 \left(w_1+\rmi w_2\right)^2 +9-12\cos4\chi \right)\overline \Sigma \right) \w\dd \chi \Bigg]\,,
\end{aligned}
\end{equation}
where $\Sigma$ is defined in \eqref{eq:sigdef}. The behavior of $C_2$ above at the locus \eqref{eq:l1sing} is incompatible with a D1-brane. Therefore we conclude that there is no candidate brane interpretation of the singularity at $\lambda=1$. To complete our analysis of the fluxes we note that the five-form flux in \eqref{eq:F5full} does not diverge in the limit $t\to 1$. Note that the function $V$ in \eqref{eq:VWdef} has additional singularities at the locus \eqref{eq:l1sing}. This in turn leads to a more singular behavior of the metric \eqref{eq:10dmetl1}. This behavior is compatible with the point-like singularities exhibited by the metric in \eqref{Eq:zoomedmetric} due to the delta-function singularities of the function $k(\alpha)$ in \eqref{eq:Hkdef}.

Some additional evidence for the peculiar nature of the naked singularity for $|\lambda|=1$ can be found by studying D3-brane probes in the full type IIB supergravity background in \eqref{eq:Kdef}-\eqref{eq:F5full}. The probe action for a D3-brane is
\begin{equation}
S_{\text{D3}} = -\f{2\pi}{(2\pi\ell_s)^4} \int  \left[\dd^4\sigma\,\sqrt{|P [g_{MN}]|} + P[C_4]\right]\,,
\end{equation}
where $\sigma_{0,1,2,3}$ are coordinates on the D3-brane world-volume and $P[\ldots]$ indicate a pullback of the metric or the 4-form RR field.
We choose the world-volume of the probe D3-brane to coincide with the  four-directions, $x_{0,1,2,3}$, that span the boundary of AdS$_5$. The embedding of the brane in the radial direction and the angles of $S^5$ is then specified by the functions $T(\sigma)$ and $\Theta^{m}(\sigma)$ for $m=1,\ldots, 5$. For static brane configurations one has to put all spacetime derivatives of the scalar functions $T(\sigma)$ and $\Theta^{m}(\sigma)$ to zero. The results is the following effective potential for these scalar functions   
\begin{equation}\label{eq:VD3def}
V_{\text{D3}} = \frac{2\pi\big( (K_1 K_2-K_3^2)^{1/2} -(1-T^2)(1-\lambda^2 T^8)\big)}{g_s (2\pi \ell_s g)^4 T^4 ( 1-\lambda^2 T^6 )^{1/3}}\,.
\end{equation}
In this expression, with a slight abuse of notation, we have used the functions $K_{1,2,3}$ in \eqref{eq:Kdef} with $T,\Theta^{1},\Theta^2$ in place of $t,\alpha,\chi$. Note that this effective potential is compatible with the $\SO(3)$ invariance of the background in \eqref{eq:Kdef}-\eqref{eq:F5full}.

The potential in \eqref{eq:VD3def} can be extremized with respect to the scalars $T,\Theta^{1},\Theta^2$ only for $\lambda=\pm1$. In addition we find that the extremum is located at $t=1$ and the locus in \eqref{eq:l1sing}.\footnote{For $\lambda=-1$ the locus in \eqref{eq:l1sing} is slighlty modified to $\alpha = n \pi/2$  for $n\in \mathbf{Z}$.} Thus we conclude that a static probe D3-brane minimizes its energy on this locus. Evaluating the full D3-brane probe action on this locus we find that it vanishes precisely. This implies that probe D3-branes become tensionless at the naked singularity for $\lambda=\pm1$. We interpret this as extra evidence that for $|\lambda|=1$ the supergravity approximation breaks down near the naked singularity and one cannot interpret it in terms of explicit brane sources.

It is worth pointing out that our analysis is similar to the calculations in \cite{Buchel:2000cn,Johnson:1999qt} where supergravity solutions dual to non-conformal $\N=2$ SYM theories were studied with peculiar naked singularities that cannot be attributed to branes. While our gauge theory setup has only $\N=1$ supersymmetry it might be possible to leverage the enhan\c{c}on mechanism of \cite{Buchel:2000cn,Johnson:1999qt}, which is associated with tensionless branes, to understand the naked singularity with $|\lambda|=1$.

%%%%%%%%%%%%%%%%
\section{Discussion}
\label{sec:discussion}
%%%%%%%%%%%%%%%%

After this detailed analysis of the type IIB uplift of the GPPZ solutions constructed in \cite{Bobev:2018eer,Petrini:2018pjk} it is worthwhile to discuss the interpretation of our results and their relation to the physics in the dual gauge theory. To this end it is important to emphasize that the supergravity solutions at hand have an $\SO(3)\times \mathbf{Z}_2$ symmetry. The $\SO(3)$ invariance is a simple manifestation of the $\SO(3)$ flavor symmetry in the $\mathcal{N}=1^{*}$ theory with three equal masses and arises from the following breaking of the $\SU(4)$ R-symmetry of $\mathcal{N}=4$ SYM
\be\label{eq:su4so3}
\SU(4)\to \SU(3)\times \U(1)_r \to \SO(3)\,.
\ee
The $\mathbf{Z}_2$ invariance is more subtle. To understand it recall that the protected operators in $\mathcal{N}=4$ SYM in the planar limit enjoy an extra $\U(1)_S$ symmetry \cite{Intriligator:1998ig}. This $\U(1)_S$ is the compact subgroup of the $\SL(2,\mathbf{R})$ duality group of planar $\mathcal{N}=4$ SYM. The $\mathbf{Z}_2$ group under which our solutions are invariant is a subgroup of $\U(1)_Y = \text{diag} (\U(1)_r \times \U(1)_S)$.  To identify candidate supersymmetric vacua of the $\mathcal{N}=1^{*}$ theory dual to our supergravity solutions we have to focus on vacua which are invariant under this $\mathbf{Z}_2$ action. It is rather unusual to impose an invariance under a subgroup of S-duality on a vacuum of a gauge theory and perhaps this feature, imposed on us by supergravity, should be attributed to the large $N$ limit in the gauge theory. Interestingly, there is a massive vacuum of $\mathcal{N}=1$ which is invariant under the $\mathbf{Z}_2$ action. It exists whenever $N=D^2$ for some integer $D$ and was discussed around \eqref{selfdualvac}. It may be tempting to speculate that for some value of $\lambda$ the GPPZ solution is dual to this massive vacuum. This interpretation is however problematic. On one hand using \eqref{eq:chgvev} we find that the chiral condensate in the selfdual vacuum \eqref{selfdualvac} is non-zero for general choices of the function $A(\tau,N)$. For vanishing $A(\tau,N)$ the IR superpotential as well as the gaugino condensate itself vanishes. On the other hand the analysis in Section~\ref{subsec:chiralcond} shows that the supergravity solutions in the four-scalar model yield zero value for the chiral condensate and gaugino condensate proportional to $\lambda$. We have arrived at this apparent contradiction by using the IR superpotential in \eqref{IRsuperpot} and the Gubser criterion on the supergravity side \cite{Gubser:2000nd}. 

%While a true sceptic may question this line of reasoning, 
We believe that the arguments above point to the fact that the GPPZ solutions with $-1< \lambda<1$ are simply not dual to massive vacua of the $\mathcal{N}=1^{*}$ theory. The near-singularity analysis for these values of $\lambda$ clearly indicate the presence of polarized five-branes in the geometry and one can attribute the singularity to the smearing of the five-branes along the ring parametrized by the coordinate $\alpha$. From this perspective, the singularity we observe in the GPPZ solution is no more unphysical than the singularities of the ${\cal N}=4$ Coulomb branch solutions in \cite{Freedman:1999gk}. We are therefore led to the conclusion that the GPPZ solutions with $-1< \lambda< 1$ are dual to a set of Coulomb vacua of $\mathcal{N}=1^{*}$ invariant under the $\mathbf{Z}_2$ action discussed above. This is not in contradiction with any of the field theory results of \cite{Dorey:1999sj,Dorey:2000fc,Polchinski:2000uf,Aharony:2000nt}. There are additional arguments in favour of our conclusion. First we note that in \cite{Polchinski:2000uf}, it was argued that a massive vacuum of $\mathcal{N}=1^{*}$ leads to a single stack of $(p,q)$ five-branes, whereas Coulomb vacua feature multiple stacks at different values of the AdS radial coordinate. The arguments of \cite{Polchinski:2000uf}  are based on the map between the classical F-term equations in the gauge theory and the Myers polarization equations for D3-branes polarizing to five-branes when immersed in flux backgrounds. In the gauge theory, the massive vacua are characterized by the fact that the sum in \eqref{eq:repsum} contains only one term. In the polarization picture of Myers this corresponds exactly to the case where the D3-branes polarize into a single stack. As soon as the sum in \eqref{eq:repsum} contains more than one term, the unbroken gauge group contains at least one unbroken $\U(1)$ and the IR physics is dominated by the dynamics of free photons. For the GPPZ backround with $-1< \lambda < 1$, we do not see a single stack of five-branes. On the contrary we find a continuous distribution of them. In fact the function $k(\alpha)$ in \eqref{eq:Hkdef} controls the tension of five-branes as a function of the angle $\alpha$ and can perhaps be given the interpretation of the $d k_d$ which appears in \eqref{eq:repsum}. At large $N$ the equation \eqref{eq:repsum} takes the form
\be\label{largeNrepsum}
\int_0^\infty x\, k_x\,\dd x =1\,,
\ee
where $x$ is the continuous analog of $d$ in \eqref{eq:repsum}. The condition \eqref{eq:int} satisfied by the function $k(\alpha)$ is indeed very reminiscent of \eqref{largeNrepsum} when $x  \sim \tan\alpha$. This mapping of $x$ to $\alpha$ is supported by the locking of the $(p,q)$ charges of the fivebranes to the coordinate $\alpha$ in our geometry.  At $\alpha=\pi/2$, which should correspond to a very large $\SU(2)$ representation, we have pure D5-branes which, according to \cite{Polchinski:2000uf}, are dual to the Higgs vacuum. On the other hand at $\alpha=0$ we expect the trivial representation of $\SU(2)$ and we find pure NS5-branes in the geometry, in line with the arguments of \cite{Polchinski:2000uf}. By this argument the vacuum described by a GPPZ solution with $-1< \lambda < 1$ corresponds to a vacuum of the theory labelled by $k_d$ which is almost ``continuous'' as a function of the dimension of the $\SU(2)$ representation $d$. Certainly the sum in \eqref{eq:repsum} contains more than one term in such a vacuum which is the case for any Coulomb vacuum. Finally, we emphasize that the behaviour of the probe string solutions we studied in detail also supports this conclusion. As explained in Section~\ref{Sec:LineOperators}, in a massive vacuum some probe $(m,n)$ string would have an on-shell action that grows linearly with the quark separation $L_{q\bar q}$. We do not find such a behavior precisely due to the continuous distribution of $(p,q)$ five-branes in the IR. 

The singularity of the GPPZ solutions with $|\lambda| = 1$ is more severe and does not seem to admit an interpretation in terms of explicit D-brane sources. The most direct evidence of this is that, as discussed in Section~\ref{subsec:lambdaequal1}, probe D3-branes become tensionless near the singularity. Given this, one might conclude that the supergravity solutions with $|\lambda| = 1$ are unphysical and thus there is no vacuum of the planar $\mathcal{N}=1^{*}$ theory with vanishing chiral condensate and a value for the gaugino bilinear vev as in  \eqref{eq:gauginovevhol}. We believe that this conclusion is erroneous. In the context of holography a natural way to excise a naked singularity of the type encountered here is to introduce an IR cutoff for the dual gauge theory. Putting the field theory at finite temperature is a natural physical choice. Indeed, this was studied for the $\mathcal{N}=1^{*}$ theory in \cite{Freedman:2000xb}, see also \cite{Bena:2018vtu} for recent work. While this is certainly one way to remedy the naked singularity of the GPPZ solutions, the lack of supersymmetry complicates the analysis of this set-up significantly. Fortunately placing the $\mathcal{N}=1^{*}$ theory on $S^4$ provides an alternative IR regulator compatible with supersymmetry. This setup was studied in detail in \cite{Bobev:2016nua} where supergravity backgrounds dual to $\mathcal{N}=1^{*}$ on $S^4$ were found as solutions of the four-scalar model in Section~\ref{sec:5Dsugra}. The solutions of \cite{Bobev:2016nua} are constructed for a fixed radius, $\mathcal{R}$, of the $S^4$ and are completely smooth. The regularity condition in the IR of the geometry translates into a precise relation between the gaugino condensate and the mass parameter for every value of $\mathcal{R}$. In Appendix~\ref{app:S4} we show that in the limit of large $\mathcal{R}$, the regular solutions of \cite{Bobev:2016nua} approach the $\lambda=1$ GPPZ solution. Since the supergravity solutions with an $S^4$ boundary are always regular, even at arbitrarily large $\mathcal{R}$, we conclude that the value $\lambda=1$ corresponds to a physical vacuum of the gauge theory. It is useful to employ an analogy with the $\mathcal{N}=2^{*}$ SYM theory. The holographic dual of $\mathcal{N}=2^{*}$ on $S^4$ is constructed in \cite{Bobev:2013cja} and we have checked explicitly that in the large $\mathcal{R}$ limit this solution asymptotes to the solution in \cite{Pilch:2000ue} with $\gamma=0$. The significance of this is that the supergravity solutions in \cite{Pilch:2000ue} are dual to $\mathcal{N}=2^{*}$ on flat space and $\gamma$ is the direct analog of the parameter $\lambda$ in $\mathcal{N}=1^{*}$. Moreover, the Gubser criterion for acceptable naked singularities restricts the range of $\gamma$ to be $\gamma\leq0$. It was shown in \cite{Buchel:2000cn} that probe D3-branes in the $\mathcal{N}=2^{*}$ solutions of \cite{Pilch:2000ue} become tensionless precisely at $\gamma=0$ which is again similar to what we find here for the $\lambda=1$ solution. The fact that the $\lambda=1$ vacuum of $\mathcal{N}=1^{*}$ is preferred by the $S^4$ IR regulator suggests that it is one of the massive vacua of the theory. This conjecture is also compatible with the fact that for $\lambda\to 1$ the function $k(\alpha)$ is peaked at four points on the $\alpha$-circle. Thus we speculate that probe strings show a qualitatively different behaviour in the $|\lambda|=1$ vacua. Now an arbitrary probe string would be unable to move along the $\alpha$ coordinate and  find a bound state with a five-brane with zero tension. This, combined with the connection between \eqref{largeNrepsum} and \eqref{eq:repsum}, suggests that the $\lambda\to 1$ has a few or even only one term in the sum and is thus massive.  The analysis above strongly suggests that the value $\lambda=1$ leads to a physical vacuum of the planar $\mathcal{N}=1^{*}$ theory. We believe that the same conclusion holds for $\lambda=-1$ but have less evidence to support this claim since there are no $S^4$ supergravity solutions which lead to this value of lambda in the large $\mathcal{R}$ limit. To shed more light on these questions it is very important to understand the vacua of $\mathcal{N}=1^{*}$ corresponding to $\lambda  =\pm 1$ using field theory methods.

Excising a naked singularity by introducing an explicit IR cutoff may not be the only mechanism to find regular supergravity solutions with $|\lambda|=1$. It is natural to wonder whether string theory provides some other mechanism to repair the singular GPPZ solutions above. The prototypical example in this context is the type IIB supergravity solution of Klebanov and Strassler \cite{Klebanov:2000hb}, which provides an explicit resolution  of the Klebanov-Tseytlin solution \cite{Klebanov:2000nc} through a geometric transition. Looking for such regular solutions in the context of $\mathcal{N}=1^{*}$ should be done directly in type IIB supergravity. Due to the small isometry group, the supergravity BPS equations reduce to a system of nonlinear partial differential equations in three variables. Finding explicit solutions of this system of equations is a daunting task. Moreover, there is no clear evidence, either from field theory or from supergravity, that regular supergravity solutions should exist. It will certainly be very interesting to settle the question about the existence of regular supergravity solutions dual to some vacua of the $\mathcal{N}=1^{*}$ theory.

Our analysis has shed new light on the holographic description of the $\mathcal{N}=1^{*}$ SYM theory with equal mass parameters. One important simplifying assumption which allowed us to make progress is that we studied configurations invariant under the $\mathbf{Z}_2$ group discussed around equation \eqref{eq:su4so3}. It is possible to relax this assumption and study a more general holographic setup with only $\SO(3)$ invariance. To this end one should employ the eight-scalar $\SO(3)$-invariant truncation of five-dimensional supergravity studied in \cite{Pilch:2000fu,Bianchi:2000sm}. This model may allow for new supersymmetric domain wall solutions with non-vanishing condensates for the scalar bilinear operators in the $\mathbf{20}'$ of $\SU(4)$. It will be very interesting to construct such solutions explicitly and uplift them to IIB supergravity using the approach outlined in  \cite{Bobev:2018eer,Petrini:2018pjk}. Moreover this eight-scalar model may allow for more general solutions with an $S^4$ boundary which may be relevant to the $\lambda=-1$ GPPZ background in the large $\mathcal{R}$ limit.

%%%%%%%%%%%%%%%%%%%%%%%%%%%%%%%%%%%%%
\bigskip
\bigskip
\leftline{\bf Acknowledgements}
\smallskip
\noindent We would like to thank Ofer Aharony, Marco Baggio, Iosif Bena, Oren Bergman, Anthony Charles, Ben Freivogel, Igor Klebanov, Michela Petrini, Krzysztof Pilch, Silviu Pufu, Jorge Santos, and Kostas Skenderis for useful discussions. The work of NB is supported in part by an Odysseus grant G0F9516N from the FWO. FFG is a Postdoctoral Fellow of the Research Foundation - Flanders (FWO). The work of BEN is supported by ERC grant ERC-2013-CoG 616732-HoloQosmos, and by the FWO and European Union's Horizon 2020 research and innovation program under the Marie Sk\l{}odowska-Curie grant agreement No. 665501. BEN is an FWO [PEGASUS]$^2$ Marie Sk\l{}odowska-Curie Fellow. The work of JvM is supported by a doctoral fellowship from the Research Foundation - Flanders (FWO). We are also supported by the KU Leuven C1 grant ZKD1118 C16/16/005.  NB and FFG would like to thank the Mainz Institute for Theoretical Physics (MITP) of the Cluster of Excellence PRISMA+ (Project ID 39083149) for hospitality during the final stages of this project.
 
 \appendix
 
%%%%%%%%%%%
\section{BPS equations for the four-scalar model}
\label{5DBPSeq}
%%%%%%%%%%%

The BPS equations of the four-scalar model with flat slicing can be written down compactly as in \eqref{5DBPSflat}. However, for the analysis in Section~\ref{subsec:chiralcond} it is more convenient to use the scalars fields $\left(\alpha,\varphi,\phi,\phi_4\right)$ as defined in \eqref{fieldredefinition}. In addition we use the metric function $A(r)$ in \eqref{eq:5dmetansatz} as a new radial variable so that the five-dimensional metric takes the form 
\begin{equation}
\dd s_5^2 = \frac{8}{g^2} \frac{ \cos \left( 3\phi - \phi_4 \right)\cos ^3\left( \phi+\phi_4 \right)}{\cos 4\phi + \cosh 4\alpha} \dd A^2 + \rme^{2A} \dd s_4^2\,.
\end{equation}
The BPS equations for the four scalar fields can then be written as
\begin{equation}\label{eq:BPSeq4scphi}
\begin{aligned}
\frac{\dd \left(\alpha-\varphi\right)}{\dd A} &= -\frac{\sinh 4\alpha}{\cos 4 \phi + \cosh 4\alpha}\cos^2\left( \phi+\phi_4 \right)\,,\\
 \frac{\dd \left(3\alpha + \varphi\right)}{\dd A} &=-\frac{3\sinh 4\alpha}{\cos 4 \phi + \cosh 4\alpha} \cos^2 \left( 3\phi - \phi_4 \right)\,,\\
 \frac{\dd \left( \phi + \phi_4 \right)}{\dd A} &=\frac{2 \sin \left( 3\phi-\phi_4 \right)- \sin \left( 5 \phi + \phi_4 \right) -3\cosh 4\alpha \sin \left( \phi+\phi_4 \right)}{\cos 4\phi+\cosh 4\alpha}\cos \left( \phi+\phi_4 \right)\,,\\
\frac{\dd \left(3 \phi - \phi_4 \right)}{\dd A} &=\, 3\frac{ \sin\left( \phi+\phi_4 \right) - \cosh 4\alpha \sin \left( 3\phi - \phi_4 \right)}{\cos4\phi+\cosh 4\alpha}\cos\left( 3\phi-\phi_4 \right)\,.
\end{aligned}
\end{equation}
Note that the five-dimensional dilaton, $\varphi$, does not appear on the right hand side of these equations. Therefore once a solution for the scalars $\alpha,\phi$ and $\phi_4$ is found the solution for $\varphi$ can be found by quadratures.

As discussed in Section~\ref{sec:5Dsugra} for the GPPZ solution one finds $\alpha=\varphi=0$ and only the scalars $\phi$ and $\phi_4$ are nontrivial. Here we provide some details on solutions of the equations \eqref{eq:BPSeq4scphi} with nontrivial $\alpha$. As discussed in in Section~\ref{subsec:chiralcond}  this scalar is dual to a protected scalar bilinear operator in the $\mathcal{N}=1^{*}$ theory.

Before we discuss general solutions of the equations in \eqref{eq:BPSeq4scphi} it is worth pointing out that there is a simple analytic solution of \eqref{eq:BPSeq4scphi} with $\phi=\phi_4=\varphi=0$. The BPS equation for $\alpha$ is then easily integrated to find
\begin{equation}
\frac{\dd \alpha}{\dd A} = -\tanh 4\alpha\quad \Rightarrow \quad\alpha = \frac{1}{4}\text{arcsinh}\left(c_1\,\rme^{-4A}\right)\,,
\end{equation}
where $c_1$ is a real integration constant. The five-dimensional metric reads  
\begin{equation}
\dd s_5^2 =\frac{8}{g^2} \frac{1}{1+ \sqrt{1+ c_1^2 \rme^{-8A}}} \dd A^2 +  \rme^{2A} \dd s_4^2\,.
\end{equation}
This solution exhibits a naked singularity, however one finds that the five-dimensional scalar potential in \eqref{Eq: 5D lagrangian} evaluated on the solution is 
\begin{equation}
\mathcal{P} = -\frac{3 g^2}{16}\left( 3 + \sqrt{1+c_1^2 \rme^{-8A}} \right)\,.
\end{equation}
This function is bounded above for all values of $c_1$ and therefore is acceptable according to the Gubser criterion in \cite{Gubser:2000nd}. This simple solution is one of the ``Coulomb branch solutions'' described in \cite{Freedman:1999gk}. It correspond to a supersymmetric vacuum of $\mathcal{N}=4$ SYM in which a particular operator in the $\mathbf{20}'$ acquires a vacuum expectation value.

The general BPS equations in \eqref{eq:BPSeq4scphi} do no admit regular solutions. This necessitates a careful study of the singular solutions. For similar BPS holographic RG flows it was argued in \cite{Gubser:2000nd} that the IR behavior of non-compact scalar fields should be such that they asymptote to a fixed direction in the scalar field space. Assuming that this behavior is indeed realized we can proceed and treat the equations in  \eqref{eq:BPSeq4scphi} in the neighborhood of the IR singularity.\footnote{We have confirmed through extensive numerical checks of the full non-line equations in \eqref{eq:BPSeq4scphi} that this assumption is indeed justified.} Since the scalars $\phi$ and $\phi_4$ are compact and periodic they should approach a constant value in order to realized the IR behavior discussed in  \cite{Gubser:2000nd}. The scalar $\alpha$ is non-compact and is not a priori restricted in the IR.

When the IR value of $\alpha$ is not vanishing one finds from \eqref{eq:BPSeq4scphi} that the constant values of the scalars $\phi$ and $\phi_4$ should obey
\begin{equation}
\begin{aligned}\label{Eq: constant condition in IR}
\cos \left( \phi^{(\text{IR})}+\phi_4^{(\text{IR})} \right)=0\,,\quad \text{and} \quad  \cos\left( 3\phi^{(\text{IR})}-\phi_4^{(\text{IR})} \right)=0\,,
\end{aligned}
\end{equation}
which means that $ \phi^{(\text{IR})} + \phi_4^{(\text{IR})}  = \pm \pi/2 = 3\phi^{(\text{IR})}-\phi_4^{(\text{IR})}$. Since the equations in \eqref{eq:BPSeq4scphi} are invariant under a simultaneous shift of $\pi/2$ in both $\phi$ and $\phi_4$ we can focus on the cases where $\phi^{(\text{IR})}=\phi_4^{(\text{IR})}-\pi/2=0$ (case I) and $\phi^{(\text{IR})}=\phi_4^{(\text{IR})}=\pi/4$ (case II). To see whether the potential is bounded from above one has to expand the scalars to second order in the IR
\begin{equation}   \label{Eq: IR expansions}
\begin{aligned}
	\text{case I}\,\,&\def\arraystretch{1.5}\left\{
    \begin{array}{rcl}
    \alpha & = & \,\alpha^{(\text{IR})} -\tanh 2\alpha^{(\text{IR})}\,\frac{a^2 + 3 b^2}{24}\rme^{6A}+ \, \mathcal O \left(\rme^{12 A}\right)\,,\\
    \phi + \phi_4 & = & \,\frac{\pi}{2} + a \rme^{3A}  + \mathcal O \left(\rme^{9 A}\right)\,,\\
     3\phi-\phi_4 & = & \,-\frac{\pi}{2} + b \rme^{3A} + \mathcal O \left(\rme^{9 A}\right)\,,
    \end{array}
    \right.\\
    \left.\right.\\
    \text{case II}\,\,&\def\arraystretch{1.5}\left\{
    \begin{array}{rcl}
    \alpha & = & \,\alpha^{(\text{IR})} -\coth 2\alpha^{(\text{IR})}\,\frac{a^2 + 3 b^2}{24}\rme^{6A} +  \, \mathcal O \left(\rme^{12 A}\right)\,,\\
    \phi + \phi_4 & = & \,\frac{\pi}{2} + a \rme^{3A}    + \mathcal O \left(\rme^{9 A}\right)\,,\\
    3\phi-\phi_4 & = & \,\frac{\pi}{2} + b \rme^{3A}    + \mathcal O \left(\rme^{9 A}\right)\,,
    \end{array}
    \right.
    \end{aligned}
\end{equation}
where $a,b$ and $\alpha^{(\text{IR})}$ are independent constants. 

To understand whether a given naked singularity in the IR is acceptable or not we once again employ the Gubser criterion. The acceptable singularities have an on-shell scalar potential that is bounded above. Evaluating the scalar potential for the 4-scalar model using the IR expansions for the two cases in \eqref{Eq: IR expansions} we find 
\begin{equation}
\begin{aligned}
\mathcal P^{(\text{IR})}_{\text{I}} \approx&- 3 g^2 \frac{\cosh^2 2\alpha^{(\text{IR})}}{8 a^3 b}\, \rme^{-12A} + g^2\frac{a - 3 b +2 a\cosh 4\alpha^{(\text{IR})}}{16 a^2 b}\,\rme^{-6A}\,,\\
 \mathcal P^{(\text{IR})}_{\text{II}} \approx&\, 3 g^2 \frac{\sinh^2 2\alpha^{(\text{IR})}}{8 a^3 b}\, \rme^{-12A} - g^2\frac{ b^2-5a^2 + 12 ab + 2 \left(b^2+5a^2\right)  \cosh 4\alpha^{(\text{IR})}}{32 a^3 b}\,\rme^{-6A}\,.
\end{aligned}
\end{equation}
Since the functaion $e^{-A}$ diverges in the IR we find that in case I the potential is bounded above when $\text{sign}(a)=\text{sign}(b)$. In case II the condition is $\text{sign}(a)=-\text{sign}(b)$. However, whenever these criteria are met we find that the full non-linear solution of the equations in \eqref{eq:BPSeq4scphi} is singular in the UV, i.e. for large $e^{A}$, and does not reach the asymptotically AdS$_5$ region. These singular UV solutions are depicted in the upper area of Figure~\ref{gubserplot}. We thus conclude that there are no physically acceptable singular solutions of the BPS equations in \eqref{eq:BPSeq4scphi} which have non-vanishing $\alpha$, $\phi$, and $\phi_4$.

When the scalar $\alpha$ vanishes we find that the only solutions to the BPS equations \eqref{eq:BPSeq4scphi} are the GPPZ solutions in Section~\ref{subsec:GPPZ} parametrized by the integration constant $\lambda$. 

 %%%%%%%%%%%
 \section{Method of near-singularity limits}
 \label{app:singularity}
  %%%%%%%%%%%

When discussing a ``near-singularity limit'' in a complicated geometry such as the uplifted GPPZ solution, it is important to clarify what one means.  A first concern is that in a background of many dimensions, a singularity may look differently depending on the direction in which it is approached; however, a more fundamental issue is what one means by the words ``near-singularity limit'' in the first place, as there are multiple different ways in which one might want to understand the structure of a singularity.  In this work, we are interested foremost in ten-dimensional brane physics, so the notion of ``near-singularity limit'' we use is meant to examine what the full 10-dimensional geometry looks like as the singularity is approached.\footnote{Other possible methods might include, for example, constructing a sort of ``pullback metric'' onto the singularity, which throws away those directions of spacetime which do not blow up as the singularity is approached.}  Such a limit contains enough data to discover the brane content of the singularity itself via, e.g., the Gauss law.

It is instructive to think of the metric as a $10 \times 10$ matrix in some (not necessarily orthonormal) basis $v^a$, thus the line element is written
\begin{equation}
\dd s^2 = g_{ab} v^a v^b\,.
\end{equation}
The metric tensor $g_{ab}$ and the basis $v^a$ are both functions of some coordinates $x^\mu$, and one is interested in their behavior as $x \to x_0$, which we can organize schematically in terms of some ``radial'' coordinate $r \equiv |x - x_0|$ (note that one should think of ``$x_0$'' as being a subspace of coordinate space which is not necessarily a single point; likewise, the singularity in the geometry may not be a single point but rather have some extension).  Then one should imagine expanding quantities as a series in powers of $r$, which may contain negative powers (for simplicity, we assume that the coordinates can be chosen such that only integer powers appear).

Here one has some choices to make about how to organize such a series.  As a $10 \times 10$ matrix, the metric tensor $g_{ab}$ has certain properties (namely, it is symmetric and invertible, and has $(1,9)$ signature), and in order to discuss a ten-dimensional near-singularity limit, we must retain these properties.  The only basis-independent information in $g_{ab}$ are its eigenvalues, so we must construct a limit in such a way as to track the behavior of each eigenvalue independently as $r \to 0$.

Any symmetric matrix can be diagonalized by an $\SO(n)$ rotation, so we can always write
\begin{equation}
g_{ab} = (R \Lambda R^\top)_{ab}\,,
\end{equation}
where $R \in \SO(10)$ (or $\SO(1,9)$; the distinction will not matter here), and $\Lambda$ is a \emph{diagonal} matrix of eigenvalues.  Since the eigenvalues of $R$ are always unitary, the singular behavior of $g_{ab}$ is now entirely contained in $\Lambda$.  It is convenient to define a new basis $\tilde v^a \equiv (R v)^a$, in which the line element is now diagonal:
\begin{equation}
\dd s^2 = \Lambda_{ab} \tilde v^a \tilde v^b = \sum_a \Lambda_{aa} (\tilde v^a)^2\,.
\end{equation}
For further convenience, we could also take $R \in \SL(10)$, which makes it somewhat easier to deal with basis-vector expressions like $\big( \sigma_3 + P(t,\alpha,\chi) \dd \alpha + Q(t,\alpha,\chi) \dd \chi \big)$, which contain linear combinations of other basis vectors.

Next one simply takes the lowest-order expansion in $r$ of each of the eigenvalues in $\Lambda$, combined with the lowest-order expansion of the rotation matrix $R$.  Since $R$ is unitary, its lowest-order expansion is always finite, of order $r^0$.  $\Lambda$ becomes a diagonal matrix of expressions with different powers of $r$:
\begin{equation}
\Lambda = \begin{pmatrix}
r^{n_0} f_0(x_\parallel) & & \\
& \ddots & \\
& & r^{n_9} f_9(x_\parallel)\,,
\end{pmatrix},
\end{equation}
where $x_\parallel$ are the coordinates parallel to the singularity (i.e. transverse to $r$).  Although each eigenvalue in $\Lambda$ may have a different order in $r$, they are each oriented along a different direction in spacetime (given by the orthogonal vectors $\tilde v^a$), and thus do not ``mix'' in a way that would allow the lower powers of $r$ to wash out the higher ones.\footnote{While this description may sound contrived at first, we point out that the result is exactly what one would get if one took a \emph{numerical} matrix and chose to truncate each of its entries to its first $n$ significant digits.}  This method of separately keeping the lowest order eigenvalues thus gives a basis-independent way of determining the local 10-dimensional geometry in the vicinity of the singularity, and in particular allows one to extract expressions which resemble D-brane metrics of the type
\begin{equation}
\dd s^2 = H^{-1/4} \dd s_6^2 + H^{3/4} \dd s_4^2\,,
\end{equation}
where the ``harmonic function'' $H$ appears with different powers in front of different parts of the metric.  Thus it is appropriate for obtaining the 10-dimensional physics of the singularity.  We note also that this is precisely the type of near-singularity limit considered elsewhere in the literature, such as in \cite{2015arXiv150304128B,Niehoff:2016gbi}.

In order to do further calculations with such a limit, one must take care to be consistent.  First, since all quantities appear only to lowest order in $r$, there is no notion of curvature as that requires two derivatives.  In order to discuss the limits of the $p$-form potentials and field strengths, it is helpful to work in the \emph{orthonormal} basis
\begin{equation}
e^a \equiv \sqrt{\Lambda_{aa}} \, \tilde v^a, \qquad \text{no sum over } a\,.
\end{equation}
One can then consistently write sums of different $e^a$ together, and their wedge products, and determine the lowest-order term in such a sum, as it is precisely the term with the lowest power of $r$ out front.  Taking care with the error terms representing the next order of $r$, one should find that all equations of motion and relations such as $F_{p+1} = \dd C_p$ are formally consistent, although many will simply vanish identically.

This method was used to obtain the near-singularity expressions given in Section~\ref{sec:10Dsolution}.  The axion-dilaton matrix, since it is a symmetric matrix, can be dealt with in the same way.

%%%%%%%%%%%
 \section{The large-radius limit of ${\cal N}=1^*$ on $S^4$}
 \label{app:S4}
 %%%%%%%%%%%

In this appendix we show how the large radius limit of the solutions obtained in \cite{Bobev:2016nua} reduce to a Euclidean version of the GPPZ solution with $\lambda =1$. In \cite{Bobev:2016nua}, it was shown that to construct supersymmetric Euclidean domain wall solutions with $S^4$ slices requires all four scalars discussed in Section \ref{sec:5Dsugra} to be turned on. Two complications arise when trying to find such spherical domain wall solutions of five-dimensional supergravity. First, the Lorentzian supergravity model must be analytically continued to Euclidean signature. In practice this means that the scalars $z_{1,2}$ and their complex conjugates $\bar z_{1,2}$ must be treated as independent scalar fields. We replace all conjugate scalars $\bar z_i$ with the symbol $\tilde z_i$ to emphasize this distinction. Second, the BPS equations for the metric and scalar fields of the supergravity theory have to be modified. To be more explicit we adopt the following metric for a spherical domain wall solution
\be\label{eq:S4dwansatz}
\dd s_5^2 = \dd r^2 + {\cal R}^2\e^{2A}\dd \Omega_4^2\,,
\ee
where $\dd \Omega_4^2$ denotes the round metric on $S^4$ with unit radius. Notice that we have introduced an explicit parameter ${\cal R}$ which can be formally thought of as the radius of the $S^4$. This parameter was omitted in the discussion of \cite{Bobev:2016nua} since it can be rescaled away by redefining the metric function $A$. Nevertheless, we find it instructive to keep it explicit in order to explore the large radius limit of $S^4$ more carefully. 

The BPS equations for the model in Section~\ref{sec:5Dsugra} with metric \eqref{eq:S4dwansatz} are
\bea
(A')^2 &=&{\cal R}^{-2}\e^{-2A}+ \f{4}{9} \e^{\cal K}  {\cal W}\widetilde{\cal W}\,,\label{ABPS}\\
(A'+s_1{\cal R}^{-1}\e^{-A})(z^i)' &=& -\f{2}{3}\e^{\cal K} {\cal W}{\cal K}^{i\tilde{\jmath}}D_{\tilde{\jmath}}\widetilde{\cal W}\,,\label{ABPS2}\\
(A'-s_1{\cal R}^{-1}\e^{-A})(\tilde{z}^{\tilde{\imath}})' &=& -\f{2}{3}\e^{\cal K} \widetilde{\cal W}{\cal K}^{\tilde{\imath}j}D_{j}{\cal W}\,.\label{ABPS3}
\eea
Here prime denotes a derivative with respect to $r$ and the parameter $s_1=\pm1$ reflects a choice of a conformal Killing spinor on $S^4$. The superpotential and K{\"a}hler potential are the same as in Section~\ref{sec:5Dsugra} but now with ${\bar z}_i$ replaced by ${\tilde z}_i$. The conjugate superpotential $\overline{\cal W}$ has similarly been replaced by $\widetilde{\cal W}$. It is easy to demonstrate that all equations of motion are satisfied as a result of these BPS equations for either choice of $s_1$. The value in keeping the parameter $\mathcal{R}$ explicit is that the BPS equations with flat slicing, i.e. domain walls with metric \eqref{eq:S4dwansatz} with $\mathbf{R}^4$ instead of $S^4$, can be obtained directly from the equations in \eqref{ABPS}-\eqref{ABPS3} by taking the limit ${\cal R} \to \infty$.

A simple solution of the equations in \eqref{ABPS}-\eqref{ABPS3} is AdS$_5$ (or rather $\mathbf{H}_5$) given by setting $z_{1,2}=\tilde{z}_{1,2}=0$ and $A = \log(\frac{4}{g\mathcal{R}} \sinh(gr/4))$. We can expand the BPS equations around this AdS vacuum and find a perturbative solution with non-trivial scalars in the UV, i.e. in the large $r$ limit. To this end we use the change of variables introduced in \eqref{fieldredefinition} combined with a similar transformation for $\tilde{z}_{1,2}$ 
\be\label{fieldredefinitiontilde}\begin{split}
\tilde z_1&= \tanh\f12\left(3\alpha+\varphi+3i\phi-i\phi_4\right)\,,\\
\tilde z_2&=\tanh\f12\left(\alpha-\varphi+i\phi+i\phi_4\right)\,.
\end{split}
\ee
Note that in general the scalars $\alpha$, $\varphi$, $\phi$ and $\phi_4$ should be treated as complex scalar fields in Euclidean signature. In terms of these variables the leading order UV expansion takes the form
\be
\begin{split}\label{eq:UVexpS4}
\phi &= \hat m \epsilon^{1/2}+ {\cal O}(\epsilon^{3/2}\log\epsilon)\,,\\
\phi_4 &= w \epsilon^{3/2}+ {\cal O}(\epsilon\log\epsilon)^2\,,\\
\alpha &= \left(v+\f{s_1 \hat m \rmi \log \epsilon}{g {\cal R}}\right) \epsilon+ {\cal O}(\epsilon\log\epsilon)^2\,,\\
\varphi &= \varphi_0 +  {\cal O}(\epsilon\log\epsilon)^2\,,\\
A &= -\f{1}{2}\log\epsilon - \left(\f{\hat m^2}{2} -\f{1}{g^2{\cal R}^2}\right)\epsilon + {\cal O}(\epsilon\log\epsilon)^2\,.
\end{split}
\ee
Notice that our expansion parameters are not the same as the ones in \cite{Bobev:2016nua}. More precisely
\be
w_\text{BEKOP} = -\rmi w\,,\quad \mu_\text{BEKOP} = \rmi\hat m\,,\quad s_\text{BEKOP} = \tanh(\varphi_0/2)\,.
\ee
In \cite{Bobev:2016nua} it was shown that there are solutions of the BPS equations \eqref{ABPS}-\eqref{ABPS3} for which the metric in \eqref{eq:S4dwansatz} caps off smoothly at some value $r=r_*$, i.e. $e^{2A}$ approaches $(r-r_*)^2$. This IR regularity condition for the spherical domain wall solution implies a relation between the UV parameters $w$ and $v$ and the mass parameter $\hat m$ in \eqref{eq:UVexpS4}. In particular, the numerical results of \cite{Bobev:2016nua} strongly suggest the relation
\be\label{wofmu}
w = 2\hat m^3\,.
\ee
This relation has been derived recently by a perturbative method in \cite{Kim:2019rwd} and we have furthermore verified through extensive numerical checks that the relation in \eqref{wofmu} is not dependent on the radius parameter ${\cal R}$. On the other hand the relation between $v$ and $\hat m$ is sensitive to the value of  ${\cal R}$ in such a way that for large $\mathcal{R}$ one finds $v\sim 1/g {\cal R}$. We therefore conclude that in the large ${\cal R}$ limit, in which the sphere is approximately $\mathbf{R}^4$, the regularity of the supergravity domain wall solution fixes  $w = 2\hat m^3$. Using the relation in \eqref{UVexpflat} we find that this is equivalent to
\be
\lambda = 1\,.
\ee
This strongly suggest that the $\lambda = 1$ GPPZ solution and its ten-dimensional uplift can be regularized by using $S^4$ as a ``supersymmetric IR cutoff''.

%%%%%%%%%%%%%
\section{Coordinates on $S^5$}
\label{app:S5}
%%%%%%%%%%%%%

To make the $\SO(3)$ isometry of the uplifted GPPZ solution manifest one has to choose appropriate coordinates on $S^5$. Our choice of coordinates differs from the one used made in \cite{Pilch:2000fu} and \cite{Petrini:2018pjk}. Here we provide an explicit map between the two sets of coordinates. The coordinates used in \cite{Pilch:2000fu,Petrini:2018pjk} are denoted with a tilde, while the ones used in this paper as well as \cite{Bobev:2018eer} are without a tilde.
 
Following \cite{Pilch:2000fu} in appendix C of \cite{Petrini:2018pjk} an explicit choice for the coordinates on the unit radius $S^5$ was made by embedding it in $\mathbf{R}^6$ with flat coordinates
 \begin{equation}
  \tilde y=\left(\tilde u_1,\tilde u_2,\tilde u_3,\tilde v_1,\tilde v_2,\tilde v_3\right)\,,
 \end{equation}
 which obey $\tilde{u}.\tilde{u}+\tilde{v}.\tilde{v}=1$. Solutions of this equation can be parametrized by a generic $\SO(3)$ rotation matrix $\tilde{R}$ and two additional angles $\left(\theta,\phi\right)$ such that
 \begin{equation}
 \tilde u = \tilde R_{\tilde \alpha_1,\tilde \alpha_2,\tilde \alpha_3} \tilde u_0, \qquad \tilde v = \tilde R_{\tilde \alpha_1,\tilde \alpha_2,\tilde \alpha_3} \tilde v_0\,,
 \end{equation}
 where $\tilde \alpha_i$ are the Euler angles parametrizing the $\SO(3)$ rotation and 
 \begin{equation}
 \tilde u_0 = \left( 0,0,\cos \tilde \theta \right),\qquad  \tilde v_0 = \left( 0,\sin \tilde \theta \sin \tilde \phi,\sin \tilde \theta \cos \tilde \phi \right)\,.
 \end{equation}
 Choosing the $\SO(3)$ matrix to be
 \begin{equation}
 \tilde R_{\tilde \alpha_1,\tilde \alpha_2,\tilde \alpha_3} = \rme^{-\tilde \alpha_3 \tilde g_1}\rme^{\tilde \alpha_2 \tilde g_2}\rme^{-\tilde \alpha_1 \tilde g_1}
 \end{equation}
 where $\left[ \tilde g_i \right]_{jk} = -\varepsilon_{ijk}$ are the generators of $\SO(3)$, and $\varepsilon_{123} =1$ we find that the metric on the round $S^5$ of unit radius becomes
 \begin{equation}
 \dd \tilde s_{S^5}^2 = \dd \tilde\theta^2 + \cos^2 \tilde\theta \left( \tilde\sigma_1^2 + \tilde\sigma_3^2\right) + \sin^2\tilde\theta \left( \sin \tilde\phi \,\tilde\sigma_2 - \cos\tilde\phi \,\tilde\sigma_1 \right)^2 + \sin^2 \tilde\theta \left( \dd \tilde\phi + \tilde\sigma_3 \right)^2\,.
 \end{equation}
This is the metric on the round  $S^5$ used in \cite{Pilch:2000fu} and \cite{Petrini:2018pjk}. In particular the $\SO(3)$ left-invariant 1-forms take the form
\be
\begin{split}\label{eq:tsigmaidef}
\tilde{\sigma}_1 =& \cos\alpha_1 \dd\alpha_2 + \sin\alpha_1\sin\alpha_2 \dd\alpha_3~,\\
\tilde{\sigma}_2 =& \sin\alpha_1 \dd\alpha_2 - \cos\alpha_1\sin\alpha_2 \dd\alpha_3~,\\
\tilde{\sigma}_3 =& \dd\alpha_1+\cos\alpha_2~\dd\alpha_3~.
\end{split}
\ee

In this paper, as well as in \cite{Bobev:2018eer}, a similar but different choice of coordinates is made. The embedding of $S^5$ in $\mathbf{R}^6$ is given by
 \begin{equation}
  u =  R_{\xi_1,\xi_2,\xi_3}  u_0\,, \qquad  v =  R_{\xi_1,\xi_2,\xi_3}  v_0\,,
 \end{equation}
 where
 \begin{equation}
  u_0 = \left( 0,\cos \chi\cos \alpha, \sin \chi \sin \alpha \right)\,, \qquad  v_0 = \left( 0, \cos \chi \sin \alpha, - \sin\chi \cos \alpha \right)\,.
 \end{equation}
 The angles $\xi_{1,2,3}$ are Euler angles of $\SO(3)$ and lead to the left-invariant 1-forms $\sigma_i$ defined in \eqref{eq:sigmaidef}. The metric on the round $S^5$ of unit radius in this coordinate system is given in \eqref{eq:metroundS5}.

To relate the two sets of coordinates presented above one has to identify $(\tilde u_0,\tilde v_0)$ with $(u_0,v_0)$ to find the following relation
 \begin{equation}
 \cos 2\tilde\theta = \cos 2\alpha \cos 2\chi\,,\qquad \cos \tilde\phi \sin 2\tilde\theta = \cos 2\chi \sin 2\alpha\,.
 \end{equation}
 Comparing $(\tilde u,\tilde v)$ with $(u,v)$ one can also relate the the one-forms in \eqref{eq:sigmaidef} and \eqref{eq:tsigmaidef} as follows
\begin{equation}
 \begin{aligned}
 \tilde \sigma_1 &= -\frac{\sqrt{2}}{\sqrt{1+\cos 2\alpha \cos 2\chi}}\left(\sin \alpha \sin\chi \,\sigma_1 + \cos \alpha \cos \chi \,\sigma_2\right)\,,\\
 \tilde \sigma_2 &= -\frac{\sqrt{2}}{\sqrt{1+\cos 2\alpha \cos 2\chi}}\left(\sin \alpha \sin\chi \,\sigma_2 - \cos \alpha \cos \chi \,\sigma_1\right)\,,\\
 \tilde \sigma_3 &= \sigma_3 + \frac{\sin 2\alpha\, \dd \chi + \sin 2\chi\, \dd \alpha}{1+\cos 2\alpha \cos 2\chi}\,.
 \end{aligned}
\end{equation}
 This provides the complete map between the coordinates used in  \cite{Pilch:2000fu} and \cite{Petrini:2018pjk} and the ones in this work (as well as \cite{Bobev:2018eer}).

\bibliography{N1star}

\providecommand{\href}[2]{#2}\begingroup\raggedright\begin{thebibliography}{10}

\bibitem{Klebanov:2000hb}
I.~R. Klebanov and M.~J. Strassler, ``{Supergravity and a confining gauge
  theory: Duality cascades and chi SB resolution of naked singularities},''
  \href{http://dx.doi.org/10.1088/1126-6708/2000/08/052}{{\em JHEP} {\bfseries
  08} (2000) 052},
\href{http://arxiv.org/abs/hep-th/0007191}{{\ttfamily arXiv:hep-th/0007191
  [hep-th]}}.
%%CITATION = HEP-TH/0007191;%%.

\bibitem{Maldacena:2000yy}
J.~M. Maldacena and C.~Nunez, ``{Towards the large N limit of pure N=1
  superYang-Mills},'' \href{http://dx.doi.org/10.1103/PhysRevLett.86.588}{{\em
  Phys. Rev. Lett.} {\bfseries 86} (2001) 588--591},
\href{http://arxiv.org/abs/hep-th/0008001}{{\ttfamily arXiv:hep-th/0008001
  [hep-th]}}.
%%CITATION = HEP-TH/0008001;%%.

\bibitem{Aharony:2000pp}
O.~Aharony, ``{A Note on the holographic interpretation of string theory
  backgrounds with varying flux},''
  \href{http://dx.doi.org/10.1088/1126-6708/2001/03/012}{{\em JHEP} {\bfseries
  03} (2001) 012},
\href{http://arxiv.org/abs/hep-th/0101013}{{\ttfamily arXiv:hep-th/0101013
  [hep-th]}}.
%%CITATION = HEP-TH/0101013;%%.

\bibitem{Gubser:2004qj}
S.~S. Gubser, C.~P. Herzog, and I.~R. Klebanov, ``{Symmetry breaking and
  axionic strings in the warped deformed conifold},''
  \href{http://dx.doi.org/10.1088/1126-6708/2004/09/036}{{\em JHEP} {\bfseries
  09} (2004) 036},
\href{http://arxiv.org/abs/hep-th/0405282}{{\ttfamily arXiv:hep-th/0405282
  [hep-th]}}.
%%CITATION = HEP-TH/0405282;%%.

\bibitem{Girardello:1999bd}
L.~Girardello, M.~Petrini, M.~Porrati, and A.~Zaffaroni, ``{The Supergravity
  dual of N=1 superYang-Mills theory},''
  \href{http://dx.doi.org/10.1016/S0550-3213(99)00764-6}{{\em Nucl. Phys.}
  {\bfseries B569} (2000) 451--469},
\href{http://arxiv.org/abs/hep-th/9909047}{{\ttfamily arXiv:hep-th/9909047
  [hep-th]}}.
%%CITATION = HEP-TH/9909047;%%.

\bibitem{Polchinski:2000uf}
J.~Polchinski and M.~J. Strassler, ``{The String dual of a confining
  four-dimensional gauge theory},''
\href{http://arxiv.org/abs/hep-th/0003136}{{\ttfamily arXiv:hep-th/0003136
  [hep-th]}}.
%%CITATION = HEP-TH/0003136;%%.

\bibitem{Pilch:2000fu}
K.~Pilch and N.~P. Warner, ``{N=1 supersymmetric renormalization group flows
  from IIB supergravity},''
  \href{http://dx.doi.org/10.4310/ATMP.2000.v4.n3.a5}{{\em Adv. Theor. Math.
  Phys.} {\bfseries 4} (2002) 627--677},
\href{http://arxiv.org/abs/hep-th/0006066}{{\ttfamily arXiv:hep-th/0006066
  [hep-th]}}.
%%CITATION = HEP-TH/0006066;%%.

\bibitem{Vafa:1994tf}
C.~Vafa and E.~Witten, ``{A Strong coupling test of S duality},''
  \href{http://dx.doi.org/10.1016/0550-3213(94)90097-3}{{\em Nucl. Phys.}
  {\bfseries B431} (1994) 3--77},
\href{http://arxiv.org/abs/hep-th/9408074}{{\ttfamily arXiv:hep-th/9408074
  [hep-th]}}.
%%CITATION = HEP-TH/9408074;%%.

\bibitem{Donagi:1995cf}
R.~Donagi and E.~Witten, ``{Supersymmetric Yang-Mills theory and integrable
  systems},'' \href{http://dx.doi.org/10.1016/0550-3213(95)00609-5}{{\em Nucl.
  Phys.} {\bfseries B460} (1996) 299--334},
\href{http://arxiv.org/abs/hep-th/9510101}{{\ttfamily arXiv:hep-th/9510101
  [hep-th]}}.
%%CITATION = HEP-TH/9510101;%%.

\bibitem{Dorey:1999sj}
N.~Dorey, ``{An Elliptic superpotential for softly broken N=4 supersymmetric
  Yang-Mills theory},''
  \href{http://dx.doi.org/10.1088/1126-6708/1999/07/021}{{\em JHEP} {\bfseries
  07} (1999) 021},
\href{http://arxiv.org/abs/hep-th/9906011}{{\ttfamily arXiv:hep-th/9906011
  [hep-th]}}.
%%CITATION = HEP-TH/9906011;%%.

\bibitem{Dorey:2000fc}
N.~Dorey and S.~P. Kumar, ``{Softly broken N=4 supersymmetry in the large N
  limit},'' \href{http://dx.doi.org/10.1088/1126-6708/2000/02/006}{{\em JHEP}
  {\bfseries 02} (2000) 006},
\href{http://arxiv.org/abs/hep-th/0001103}{{\ttfamily arXiv:hep-th/0001103
  [hep-th]}}.
%%CITATION = HEP-TH/0001103;%%.

\bibitem{Aharony:2000nt}
O.~Aharony, N.~Dorey, and S.~P. Kumar, ``{New modular invariance in the N=1*
  theory, operator mixings and supergravity singularities},''
  \href{http://dx.doi.org/10.1088/1126-6708/2000/06/026}{{\em JHEP} {\bfseries
  06} (2000) 026},
\href{http://arxiv.org/abs/hep-th/0006008}{{\ttfamily arXiv:hep-th/0006008
  [hep-th]}}.
%%CITATION = HEP-TH/0006008;%%.

\bibitem{Dijkgraaf:2002dh}
R.~Dijkgraaf and C.~Vafa, ``A perturbative window into non-perturbative
  physics,''
\href{http://arxiv.org/abs/hep-th/0208048}{{\ttfamily hep-th/0208048}}.
%%CITATION = HEP-TH 0208048;%%.

\bibitem{Gunaydin:1984qu}
M.~Gunaydin, L.~J. Romans, and N.~P. Warner, ``{Gauged N=8 Supergravity in
  Five-Dimensions},''
\href{http://dx.doi.org/10.1016/0370-2693(85)90361-2}{{\em Phys. Lett.}
  {\bfseries 154B} (1985) 268--274}.
%%CITATION = PHLTA,154B,268;%%.

\bibitem{Pernici:1985ju}
M.~Pernici, K.~Pilch, and P.~van Nieuwenhuizen, ``{Gauged N=8 D=5
  Supergravity},''
\href{http://dx.doi.org/10.1016/0550-3213(85)90645-5}{{\em Nucl. Phys.}
  {\bfseries B259} (1985) 460}.
%%CITATION = NUPHA,B259,460;%%.

\bibitem{Gunaydin:1985cu}
M.~Gunaydin, L.~J. Romans, and N.~P. Warner, ``{Compact and Noncompact Gauged
  Supergravity Theories in Five-Dimensions},''
\href{http://dx.doi.org/10.1016/0550-3213(86)90237-3}{{\em Nucl. Phys.}
  {\bfseries B272} (1986) 598--646}.
%%CITATION = NUPHA,B272,598;%%.

\bibitem{Myers:1999ps}
R.~C. Myers, ``{Dielectric branes},''
  \href{http://dx.doi.org/10.1088/1126-6708/1999/12/022}{{\em JHEP} {\bfseries
  12} (1999) 022},
\href{http://arxiv.org/abs/hep-th/9910053}{{\ttfamily arXiv:hep-th/9910053
  [hep-th]}}.
%%CITATION = HEP-TH/9910053;%%.

\bibitem{Lee:2014mla}
K.~Lee, C.~Strickland-Constable, and D.~Waldram, ``{Spheres, generalised
  parallelisability and consistent truncations},''
  \href{http://dx.doi.org/10.1002/prop.201700048}{{\em Fortsch. Phys.}
  {\bfseries 65} no.~10-11, (2017) 1700048},
\href{http://arxiv.org/abs/1401.3360}{{\ttfamily arXiv:1401.3360 [hep-th]}}.
%%CITATION = ARXIV:1401.3360;%%.

\bibitem{Baguet:2015sma}
A.~Baguet, O.~Hohm, and H.~Samtleben, ``{Consistent Type IIB Reductions to
  Maximal 5D Supergravity},''
  \href{http://dx.doi.org/10.1103/PhysRevD.92.065004}{{\em Phys. Rev.}
  {\bfseries D92} no.~6, (2015) 065004},
\href{http://arxiv.org/abs/1506.01385}{{\ttfamily arXiv:1506.01385 [hep-th]}}.
%%CITATION = ARXIV:1506.01385;%%.

\bibitem{Pilch:2000ue}
K.~Pilch and N.~P. Warner, ``{N=2 supersymmetric RG flows and the IIB
  dilaton},'' \href{http://dx.doi.org/10.1016/S0550-3213(00)00656-8}{{\em Nucl.
  Phys.} {\bfseries B594} (2001) 209--228},
\href{http://arxiv.org/abs/hep-th/0004063}{{\ttfamily arXiv:hep-th/0004063
  [hep-th]}}.
%%CITATION = HEP-TH/0004063;%%.

\bibitem{Petrini:2018pjk}
M.~Petrini, H.~Samtleben, S.~Schmidt, and K.~Skenderis, ``{The 10d Uplift of
  the GPPZ Solution},''
\href{http://arxiv.org/abs/1805.01919}{{\ttfamily arXiv:1805.01919 [hep-th]}}.
%%CITATION = ARXIV:1805.01919;%%.

\bibitem{Bobev:2018eer}
N.~Bobev, F.~F. Gautason, B.~E. Niehoff, and J.~van Muiden, ``{Uplifting GPPZ:
  A Ten-dimensional Dual of $\mathcal{N}=1^{*}$},''
\href{http://arxiv.org/abs/1805.03623}{{\ttfamily arXiv:1805.03623 [hep-th]}}.
%%CITATION = ARXIV:1805.03623;%%.

\bibitem{Gubser:2000nd}
S.~S. Gubser, ``{Curvature singularities: The Good, the bad, and the naked},''
  \href{http://dx.doi.org/10.4310/ATMP.2000.v4.n3.a6}{{\em Adv. Theor. Math.
  Phys.} {\bfseries 4} (2000) 679--745},
\href{http://arxiv.org/abs/hep-th/0002160}{{\ttfamily arXiv:hep-th/0002160
  [hep-th]}}.
%%CITATION = HEP-TH/0002160;%%.

\bibitem{Maldacena:2000mw}
J.~M. Maldacena and C.~Nunez, ``{Supergravity description of field theories on
  curved manifolds and a no go theorem},''
  \href{http://dx.doi.org/10.1142/S0217751X01003935,
  10.1142/S0217751X01003937}{{\em Int. J. Mod. Phys.} {\bfseries A16} (2001)
  822--855}, \href{http://arxiv.org/abs/hep-th/0007018}{{\ttfamily
  arXiv:hep-th/0007018 [hep-th]}}.
[,182(2000)].
%%CITATION = HEP-TH/0007018;%%.

\bibitem{Bobev:2016nua}
N.~Bobev, H.~Elvang, U.~Kol, T.~Olson, and S.~S. Pufu, ``{Holography for $
  \mathcal{N} $ = 1$^{*}$ on S$^{4}$},''
  \href{http://dx.doi.org/10.1007/JHEP10(2016)095}{{\em JHEP} {\bfseries 10}
  (2016) 095},
\href{http://arxiv.org/abs/1605.00656}{{\ttfamily arXiv:1605.00656 [hep-th]}}.
%%CITATION = ARXIV:1605.00656;%%.

\bibitem{Leigh:1995ep}
R.~G. Leigh and M.~J. Strassler, ``{Exactly marginal operators and duality in
  four-dimensional $N=1$ supersymmetric gauge theory},''
  \href{http://dx.doi.org/10.1016/0550-3213(95)00261-P}{{\em Nucl. Phys.}
  {\bfseries B447} (1995) 95--136},
\href{http://arxiv.org/abs/hep-th/9503121}{{\ttfamily arXiv:hep-th/9503121}}.
%%CITATION = HEP-TH/9503121;%%.

\bibitem{Khavaev:1998fb}
A.~Khavaev, K.~Pilch, and N.~P. Warner, ``{New vacua of gauged N=8 supergravity
  in five-dimensions},''
  \href{http://dx.doi.org/10.1016/S0370-2693(00)00795-4}{{\em Phys. Lett.}
  {\bfseries B487} (2000) 14--21},
\href{http://arxiv.org/abs/hep-th/9812035}{{\ttfamily arXiv:hep-th/9812035
  [hep-th]}}.
%%CITATION = HEP-TH/9812035;%%.

\bibitem{Freedman:1999gk}
D.~Z. Freedman, S.~S. Gubser, K.~Pilch, and N.~P. Warner, ``{Continuous
  distributions of D3-branes and gauged supergravity},''
  \href{http://dx.doi.org/10.1088/1126-6708/2000/07/038}{{\em JHEP} {\bfseries
  07} (2000) 038},
\href{http://arxiv.org/abs/hep-th/9906194}{{\ttfamily arXiv:hep-th/9906194
  [hep-th]}}.
%%CITATION = HEP-TH/9906194;%%.

\bibitem{tHooft:1977nqb}
G.~'t~Hooft, ``{On the Phase Transition Towards Permanent Quark Confinement},''
\href{http://dx.doi.org/10.1016/0550-3213(78)90153-0}{{\em Nucl. Phys.}
  {\bfseries B138} (1978) 1--25}.
%%CITATION = NUPHA,B138,1;%%.

\bibitem{Montonen:1977sn}
C.~Montonen and D.~I. Olive, ``{Magnetic Monopoles as Gauge Particles?},''
\href{http://dx.doi.org/10.1016/0370-2693(77)90076-4}{{\em Phys. Lett.}
  {\bfseries 72B} (1977) 117--120}.
%%CITATION = PHLTA,72B,117;%%.

\bibitem{Skenderis:2002wp}
K.~Skenderis, ``{Lecture notes on holographic renormalization},''
  \href{http://dx.doi.org/10.1088/0264-9381/19/22/306}{{\em Class. Quant.
  Grav.} {\bfseries 19} (2002) 5849--5876},
\href{http://arxiv.org/abs/hep-th/0209067}{{\ttfamily arXiv:hep-th/0209067}}.
%%CITATION = HEP-TH/0209067;%%.

\bibitem{Bianchi:2001kw}
M.~Bianchi, D.~Z. Freedman, and K.~Skenderis, ``{Holographic
  renormalization},'' {\em Nucl. Phys.} {\bfseries B631} (2002) 159--194,
\href{http://arxiv.org/abs/hep-th/0112119}{{\ttfamily arXiv:hep-th/0112119}}.
%%CITATION = HEP-TH/0112119;%%.

\bibitem{Bianchi:2001de}
M.~Bianchi, D.~Z. Freedman, and K.~Skenderis, ``{How to go with an RG flow},''
  \href{http://dx.doi.org/10.1088/1126-6708/2001/08/041}{{\em JHEP} {\bfseries
  08} (2001) 041},
\href{http://arxiv.org/abs/hep-th/0105276}{{\ttfamily arXiv:hep-th/0105276
  [hep-th]}}.
%%CITATION = HEP-TH/0105276;%%.

\bibitem{Marolf:2000cb}
D.~Marolf, ``{Chern-Simons terms and the three notions of charge},'' in {\em
  {Quantization, gauge theory, and strings. Proceedings, International
  Conference dedicated to the memory of Professor Efim Fradkin, Moscow, Russia,
  June 5-10, 2000. Vol. 1+2}}, pp.~312--320.
\newblock 2000.
\newblock
\href{http://arxiv.org/abs/hep-th/0006117}{{\ttfamily arXiv:hep-th/0006117
  [hep-th]}}.
\newblock
%%CITATION = HEP-TH/0006117;%%.

\bibitem{Lu:1998vh}
J.~X. Lu and S.~Roy, ``{An SL(2,Z) multiplet of type IIB super five-branes},''
  \href{http://dx.doi.org/10.1016/S0370-2693(98)00435-3}{{\em Phys. Lett.}
  {\bfseries B428} (1998) 289--296},
\href{http://arxiv.org/abs/hep-th/9802080}{{\ttfamily arXiv:hep-th/9802080
  [hep-th]}}.
%%CITATION = HEP-TH/9802080;%%.

\bibitem{Tong:2002rq}
D.~Tong, ``{NS5-branes, T duality and world sheet instantons},''
  \href{http://dx.doi.org/10.1088/1126-6708/2002/07/013}{{\em JHEP} {\bfseries
  07} (2002) 013},
\href{http://arxiv.org/abs/hep-th/0204186}{{\ttfamily arXiv:hep-th/0204186
  [hep-th]}}.
%%CITATION = HEP-TH/0204186;%%.

\bibitem{Witten:1995im}
E.~Witten, ``{Bound states of strings and p-branes},''
  \href{http://dx.doi.org/10.1016/0550-3213(95)00610-9}{{\em Nucl. Phys.}
  {\bfseries B460} (1996) 335--350},
\href{http://arxiv.org/abs/hep-th/9510135}{{\ttfamily arXiv:hep-th/9510135
  [hep-th]}}.
%%CITATION = HEP-TH/9510135;%%.

\bibitem{Witten:1998zw}
E.~Witten, ``{Anti-de Sitter space, thermal phase transition, and confinement
  in gauge theories},''
  \href{http://dx.doi.org/10.4310/ATMP.1998.v2.n3.a3}{{\em Adv. Theor. Math.
  Phys.} {\bfseries 2} (1998) 505--532},
  \href{http://arxiv.org/abs/hep-th/9803131}{{\ttfamily arXiv:hep-th/9803131
  [hep-th]}}.
[,89(1998)].
%%CITATION = HEP-TH/9803131;%%.

\bibitem{Maldacena:1998im}
J.~M. Maldacena, ``{Wilson loops in large N field theories},''
  \href{http://dx.doi.org/10.1103/PhysRevLett.80.4859}{{\em Phys. Rev. Lett.}
  {\bfseries 80} (1998) 4859--4862},
\href{http://arxiv.org/abs/hep-th/9803002}{{\ttfamily arXiv:hep-th/9803002
  [hep-th]}}.
%%CITATION = HEP-TH/9803002;%%.

\bibitem{Rey:1998ik}
S.-J. Rey and J.-T. Yee, ``{Macroscopic strings as heavy quarks in large N
  gauge theory and anti-de Sitter supergravity},''
  \href{http://dx.doi.org/10.1007/s100520100799}{{\em Eur. Phys. J.} {\bfseries
  C22} (2001) 379--394},
\href{http://arxiv.org/abs/hep-th/9803001}{{\ttfamily arXiv:hep-th/9803001
  [hep-th]}}.
%%CITATION = HEP-TH/9803001;%%.

\bibitem{Drukker:1999zq}
N.~Drukker, D.~J. Gross, and H.~Ooguri, ``{Wilson loops and minimal
  surfaces},'' \href{http://dx.doi.org/10.1103/PhysRevD.60.125006}{{\em Phys.
  Rev.} {\bfseries D60} (1999) 125006},
\href{http://arxiv.org/abs/hep-th/9904191}{{\ttfamily arXiv:hep-th/9904191
  [hep-th]}}.
%%CITATION = HEP-TH/9904191;%%.

\bibitem{Zarembo:2002an}
K.~Zarembo, ``{Supersymmetric Wilson loops},''
  \href{http://dx.doi.org/10.1016/S0550-3213(02)00693-4}{{\em Nucl. Phys.}
  {\bfseries B643} (2002) 157--171},
\href{http://arxiv.org/abs/hep-th/0205160}{{\ttfamily arXiv:hep-th/0205160
  [hep-th]}}.
%%CITATION = HEP-TH/0205160;%%.

\bibitem{Buchel:2000cn}
A.~Buchel, A.~W. Peet, and J.~Polchinski, ``{Gauge dual and noncommutative
  extension of an N=2 supergravity solution},''
  \href{http://dx.doi.org/10.1103/PhysRevD.63.044009}{{\em Phys. Rev.}
  {\bfseries D63} (2001) 044009},
\href{http://arxiv.org/abs/hep-th/0008076}{{\ttfamily arXiv:hep-th/0008076
  [hep-th]}}.
%%CITATION = HEP-TH/0008076;%%.

\bibitem{Johnson:1999qt}
C.~V. Johnson, A.~W. Peet, and J.~Polchinski, ``{Gauge theory and the excision
  of repulson singularities},''
  \href{http://dx.doi.org/10.1103/PhysRevD.61.086001}{{\em Phys. Rev.}
  {\bfseries D61} (2000) 086001},
\href{http://arxiv.org/abs/hep-th/9911161}{{\ttfamily arXiv:hep-th/9911161
  [hep-th]}}.
%%CITATION = HEP-TH/9911161;%%.

\bibitem{Intriligator:1998ig}
K.~A. Intriligator, ``{Bonus symmetries of N=4 superYang-Mills correlation
  functions via AdS duality},''
  \href{http://dx.doi.org/10.1016/S0550-3213(99)00242-4}{{\em Nucl. Phys.}
  {\bfseries B551} (1999) 575--600},
\href{http://arxiv.org/abs/hep-th/9811047}{{\ttfamily arXiv:hep-th/9811047
  [hep-th]}}.
%%CITATION = HEP-TH/9811047;%%.

\bibitem{Freedman:2000xb}
D.~Z. Freedman and J.~A. Minahan, ``{Finite temperature effects in the
  supergravity dual of the N=1* gauge theory},''
  \href{http://dx.doi.org/10.1088/1126-6708/2001/01/036}{{\em JHEP} {\bfseries
  01} (2001) 036},
\href{http://arxiv.org/abs/hep-th/0007250}{{\ttfamily arXiv:hep-th/0007250
  [hep-th]}}.
%%CITATION = HEP-TH/0007250;%%.

\bibitem{Bena:2018vtu}
I.~Bena, {\'O}.~J.~C. Dias, G.~S. Hartnett, B.~E. Niehoff, and J.~E. Santos,
  ``{Holographic dual of hot Polchinski-Strassler quark-gluon plasma},''
\href{http://arxiv.org/abs/1805.06463}{{\ttfamily arXiv:1805.06463 [hep-th]}}.
%%CITATION = ARXIV:1805.06463;%%.

\bibitem{Bobev:2013cja}
N.~Bobev, H.~Elvang, D.~Z. Freedman, and S.~S. Pufu, ``{Holography for
  $\mathcal{N} = 2^*$ on $S^4$},''
  \href{http://dx.doi.org/10.1007/JHEP07(2014)001}{{\em JHEP} {\bfseries 07}
  (2014) 001},
\href{http://arxiv.org/abs/1311.1508}{{\ttfamily arXiv:1311.1508 [hep-th]}}.
%%CITATION = ARXIV:1311.1508;%%.

\bibitem{Klebanov:2000nc}
I.~R. Klebanov and A.~A. Tseytlin, ``{Gravity duals of supersymmetric SU(N) x
  SU(N+M) gauge theories},''
  \href{http://dx.doi.org/10.1016/S0550-3213(00)00206-6}{{\em Nucl. Phys.}
  {\bfseries B578} (2000) 123--138},
\href{http://arxiv.org/abs/hep-th/0002159}{{\ttfamily arXiv:hep-th/0002159
  [hep-th]}}.
%%CITATION = HEP-TH/0002159;%%.

\bibitem{Bianchi:2000sm}
M.~Bianchi, O.~DeWolfe, D.~Z. Freedman, and K.~Pilch, ``{Anatomy of two
  holographic renormalization group flows},'' {\em JHEP} {\bfseries 01} (2001)
  021,
\href{http://arxiv.org/abs/hep-th/0009156}{{\ttfamily arXiv:hep-th/0009156}}.
%%CITATION = HEP-TH/0009156;%%.

\bibitem{2015arXiv150304128B}
O.~{Biquard}, ``{M\'etriques hyperk\"ahl\'eriennes pli\'ees},'' {\em ArXiv
  e-prints} (Mar., 2015) , \href{http://arxiv.org/abs/1503.04128}{{\ttfamily
  arXiv:1503.04128 [math.DG]}}.

\bibitem{Niehoff:2016gbi}
B.~E. Niehoff and H.~S. Reall, ``{Evanescent ergosurfaces and ambipolar
  hyperk\"ahler metrics},'' {\em JHEP} {\bfseries 04} (2016) 130,
\href{http://arxiv.org/abs/1601.01898}{{\ttfamily arXiv:1601.01898 [hep-th]}}.
%%CITATION = ARXIV:1601.01898;%%.

\bibitem{Kim:2019rwd}
N.~Kim and S.-J. Kim, ``{Perturbative solutions of ${\cal N}=1^*$ holography on
  $S^4$},''
\href{http://arxiv.org/abs/1904.02038}{{\ttfamily arXiv:1904.02038 [hep-th]}}.
%%CITATION = ARXIV:1904.02038;%%.

\end{thebibliography}\endgroup
\bibliographystyle{utphys}

\end{document}